\documentclass[preprint2]{aastex}

 \usepackage[fleqn]{amsmath}
 \usepackage{graphicx}
\usepackage{dcolumn}
\usepackage{bm}

\shorttitle{A novel energy-conserving scheme for eight-dimensional
systems} \shortauthors{Hu  et al.}

\begin{document}


\title{A Novel Energy-conserving Scheme for Eight-dimensional Hamiltonian Problems}


\author{Shiyang Hu$^{1}$,  Xin Wu$^{1,2, \dag}$, Guoqing Huang$^{1}$, Enwei Liang$^{2}$}
\affil{1. Department of Physics,
Nanchang University, Nanchang 330031, People' s Republic of China \\
2. Guangxi Key Laboratory for Relativistic Astrophysics $\&$ School of Physical Science and Technology, Guangxi University,
Nanning 530004, People' s Republic of China} \email{$\dag$ xinwu@gxu.edu.cn}



\begin{abstract}

We design a novel, exact energy-conserving implicit
nonsymplectic integration method for an eight-dimensional
Hamiltonian system with four degrees of freedom. In our algorithm,
each partial derivative of the Hamiltonian with respect to one of the
phase-space variables is discretized by the average of eight
Hamiltonian difference terms. Such a discretization form is a
second-order approximation to the Hamiltonian gradient. It is
shown numerically via simulations of a Fermi-Pasta-Ulam- $\beta$ system and a
post-Newtonian conservative system of compact binaries with one
body spinning that the newly proposed method has extremely good
energy-conserving performance, compared to the Runge-Kutta;
an implicit midpoint symplectic method, and extended phase-space
explicit symplectic-like integrators. The new method is
advantageous over very long times and for large time steps
compared to the state-of-the-art Runge-Kutta method in the accuracy of
numerical solutions. Although such an energy-conserving integrator
exhibits a higher computational cost than any one of the other
three algorithms, the superior results justify its use for
satisfying some specific purposes on the preservation of energies
in numerical simulations with much longer times, e.g., obtaining a
high enough accuracy of the semimajor axis in a Keplerian problem
in the solar system or accurately grasping the frequency of a
gravitational wave from a circular orbit in a post-Newtonian
system of compact binaries. The new integrator will be potentially
applied to model time-varying external electromagnetic fields or
time-dependent spacetimes.

\end{abstract}





\section{Introduction}

Numerical integration schemes are convenient to handle many
complicated nonlinear dynamical problems, such as the motion of
relativistic charged particles in external electromagnetic fields
(Kop\'{a}\v{c}ek $\&$ Karas 2014) and general relativistic systems
of compact binaries consisting of black holes and/or neutron stars
(Thorne $\&$ Hartle 1985; Kidder 1995; Buonanno et al. 2006).
Although explicit Runge-Kutta (RK) integrators can generally give very
accurate solutions to these systems, they are
nonsymplectic/nongeometric methods that do not conserve the
invariants of motion and, therefore, cause a secular, unphysical
increase in energy errors even if these systems are conservative.
In this sense, such non-geometric methods are generally discarded.
Instead, geometric integration methods (Hairer et al. 2006) are
usually used.

Without doubt, symplectic integrators (Ruth 1983; Feng $\&$ Qin
1987; Forest 1990; Wisdom $\&$ Holman 1991; Zhong et al. 2010) are
a class of geometric integration methods. They not only preserve
symplectic structures of Hamiltonian systems but also show no
intrinsic unbounded accumulation of energy errors due to their
long-term numerical stability. Symmetric methods (Quinlan $\&$
Tremaine 1990) are similar to the symplectic integrators and do
not lead to a secular, unphysical energy drift. Extended
phase-space methods (Pihajoki 2015; Liu et al. 2016; Luo et al.
2017; Li $\&$ Wu 2017) are also explicitly symplectic-like or are 
symmetric schemes. As a point to note, the so-called conservation
of energies in these integrators does not mean that these
algorithms strictly conserve the energy integrals without any
truncation errors from a theoretical viewpoint, but that does mean that
these algorithms show no secular drift in the energy errors from a
numerical viewpoint.

Are there a class of  energy-conserving integrators that can
exactly conserve the energy integral of a conservative Hamiltonian
from the theoretical viewpoint? Yes, there are. As a path to
obtain them, the gradient of the Hamiltonian is discretized  by
use of Hamiltonian difference terms so that the Hamiltonian can be
exactly conserved numerically step by step. This conservation of
the Hamiltonian of the system is said to be energy
conservation.\footnote{In the present paper, this presentation is
only based on the general case in which the Hamiltonian represents
some form of the energy of the system. If the Hamiltonian differs
from the energy, then Hamiltonian-conserving does not mean
energy-conserving.} First-order discrete Hamiltonian gradient
schemes were constructed by Chorin et al. (1978) and Feng (1985).
The construction of such first-order Hamiltonian-conserving
methods is simple, but those of second-order
Hamiltonian-conserving schemes may be complicated. The
second-order discretization of the Hamiltonian gradient depends on
the dimension of the Hamiltonian. The second-order discrete
gradient scheme for a 4-dimensional Hamiltonian system with two
degrees of freedom can be found in some references (Qin 1987; Itoh
$\&$ Abe 1988; Wang et al. 2008; Feng $\&$ Qin 2009). The
discretization of each partial derivative of the Hamiltonian with
respect to one of the position and momentum variables is the
average of four Hamiltonian difference terms. Recently, an
extension to a 6-dimensional Hamiltonian system with three degrees
of freedom was given by Bacchini et al. (2018a). Here, each
component of the Hamiltonian gradient is replaced with the average
of six Hamiltonian difference terms. The number of such average
values adds up to six. This energy-conserving integration scheme
was used to model time-like and null geodesics in general
relativity (Bacchini et al. 2018b). These exact energy-conserving
algorithms are implicit and nonsymplectic.

It is worth pointing out that the exact conservation of energies
is important during numerical studies of  the long-term evolution
of conservative systems in astrophysics. This requirement
satisfies a physical need of conservative systems. In addition, it
is based on a need of some specific purposes in numerical
simulations with much longer times. For example, the Keplerian
energy of a pure Keplerian two-body problem in the solar system
closely depends on the orbital semimajor axis, mean motion, and
mean anomaly (Murray $\&$ Dermott 1999). The conservation of the
Keplerian energy in numerical simulations results in improving the
accuracy of the semimajor axis, eliminating the Lyapunov
instability of orbits, and suppressing the fast accumulation of
numerical errors along the in-track direction. In particular, this
leads to accurately grasping periodic motion (i.e. the orbital
frequency of a circular, spherical or quasi-spherical orbit) in a
system of two black holes (Kidder 1995) over long times -- that is,
accurately grasping the frequency of a gravitational wave emitted
from the circular, spherical, or quasi-spherical orbit. This is
because the frequency of the wave from the circular orbit doubles
the frequency of the circular orbit.

Considering the importance of conserving energy in numerical
integrations of long-term evolution of objects in astrophysics,
we will generalize the work of Bacchini et al. (2018a) and
construct a new exact energy-conserving integration method for an
eight-dimensional Hamiltonian with four degrees of freedom in the
present paper. This scheme is also based on Hamiltonian
differencing and replaces the Hamiltonian gradient with the
average of eight Hamiltonian difference terms. This discrete
average is still a second-order approximation to the Hamiltonian
gradient. This is the main aim of this paper.

The paper is organized as follows. A new energy-conserving method is
given in Section 2. Taking a Fermi-Pasta-Ulam- $\beta$ system (Skokos et al.
2008; Gerlach et al. 2012) as a tested model in Section 3, we
check the numerical performance of the new method, compared to
that of the RK method; the implicit midpoint rule (Feng 1986; Zhong
et al. 2010; Mei et al. 2013); and extended phase-space methods
(Pihajoki 2015; Liu et al. 2016; Luo et al. 2017; Li $\&$ Wu
2017). We also focus on the dynamical features of order and chaos
of some orbits in the system. Then, a post-Newtonian conservative
system of compact binaries with one body spinning (Damour et al.
2000a, 2000b, 2001; Nagar 2011) is another tested model in Section
4. Finally, our main results are concluded in Section 5. The
extended phase-space explicit methods for inseparable Hamiltonian
problems are introduced briefly in the Appendix.

\section{A new energy-conserving method}

Let us consider an $N$-dimensional Hamiltonian problem
$H(\textbf{q},\textbf{p})$, which corresponds to the following
Hamiltonian canonical equations:
\begin{eqnarray}\label{1}
\dot{\textbf{q}} &=& \frac{\partial{H}}{\partial{\textbf{p}}},\\
\dot{\textbf{p}} &=& -\frac{\partial{H}}{\partial{\textbf{q}}}.
\end{eqnarray}
From an $n$th step to an $(n+1)$th step, these derivatives have
discrete forms:
\begin{eqnarray}\label{2}
  \frac{\mathbf{q}_{n+1}-\mathbf{q}_{n}}{\Delta{t}} &=& [H(\mathbf{q}_{n},\mathbf{p}_{n+1})
  -H(\mathbf{q}_{n},\mathbf{p}_{n})] \nonumber \\
  && /(\mathbf{p}_{n+1}-\mathbf{p}_{n}), \\
  \frac{\mathbf{p}_{n+1}-\mathbf{p}_{n}}{\Delta{t}} &=&
 [H(\mathbf{q}_{n},\mathbf{p}_{n+1})-H(\mathbf{q}_{n+1},\mathbf{p}_{n+1})] \nonumber \\
  &&  /(\mathbf{q}_{n+1}-\mathbf{q}_{n}).
\end{eqnarray}
It is easy to obtain from the two equations that
\begin{equation}\label{4}
H(\mathbf{q}_{n+1},\mathbf{p}_{n+1})-H(\mathbf{q}_{n},\mathbf{p}_{n})
= 0.
\end{equation}
That means that the discretized Equations (3) and (4) do exactly
conserve the Hamiltonian. Therefore, they are an energy-conserving
scheme in the general case. In fact, the discrete gradient is only
a first-order approximation to the gradient of the Hamiltonian,
$\nabla H$. In what follows, we consider how to construct an
energy-conserving scheme with a second-order approximation to the
gradient of an 8-dimensional Hamiltonian with four degrees of
freedom.

Now, let $\mathbf{q}$ be a 4-dimensional position vector, and
$\mathbf{p}$ be a four-dimensional momentum vector, i.e.
$\textbf{q}=(q_{1},q_{2},q_{3},q_{4})$ and
$\textbf{p}=(p_{1},p_{2},p_{3},p_{4})$. $h=\Delta t$ is a time
step. For simplicity, the value of $H(q_{1n},q_{2n},q_{3n}$,
$q_{4n}$, $p_{1n},$ $p_{2n}$, $p_{3n}$, $p_{4n})$ at an $n$th step
is labeled as $H(00000000)$, and that of
$H(q_{1(n+1)},q_{2(n+1)},q_{3(n+1)}, q_{4(n+1)}, p_{1(n+1)}$,
$p_{2(n+1)},p_{3(n+1)},p_{4(n+1)})$ at an $(n+1)$th step is marked
as $H(11111111)$. In addition, we take $q_{1n}=q_{10}$,
$q_{1(n+1)}=q_{11}$, $\cdots$. Using these notations, we
discretize Equations (1) and (2) as follows:
\begin{eqnarray}\label{13}
&& \frac{q_{11}-q_{10}}{h}= \nonumber\\
&& \frac{1}{8}[\frac{H(00001000)-H(00000000)}{p_{11}-p_{10}} \nonumber\\
&& +\frac{H(00011000)-H(00010000)}{p_{11}-p_{10}} \nonumber\\
&& +\frac{H(00011001)-H(00010001)}{p_{11}-p_{10}} \nonumber\\
&& +\frac{H(00111001)-H(00110001)}{p_{11}-p_{10}} \nonumber\\
&& +\frac{H(00111011)-H(00110011)}{p_{11}-p_{10}} \nonumber\\
&& +\frac{H(01111011)-H(01110011)}{p_{11}-p_{10}} \nonumber\\
&& +\frac{H(01111111)-H(01110111)}{p_{11}-p_{10}} \nonumber\\
&& +\frac{H(11111111)-H(11110111)}{p_{11}-p_{10}}],
\end{eqnarray}
\begin{eqnarray}
&&  \frac{q_{21}-q_{20}}{h}= \nonumber\\
&& \frac{1}{8}[\frac{H(00000100)-H(00000000)}{p_{21}-p_{20}} \nonumber\\
&& +\frac{H(10000100)-H(10000000)}{p_{21}-p_{20}} \nonumber\\
&& +\frac{H(10001100)-H(10001000)}{p_{21}-p_{20}} \nonumber\\
&& +\frac{H(10011100)-H(10011000)}{p_{21}-p_{20}} \nonumber\\
&& +\frac{H(10011101)-H(10011001)}{p_{21}-p_{20}} \nonumber\\
&& +\frac{H(10111101)-H(10111001)}{p_{21}-p_{20}} \nonumber\\
&& +\frac{H(10111111)-H(10111011)}{p_{21}-p_{20}} \nonumber\\
&& +\frac{H(11111111)-H(11110111)}{p_{21}-p_{20}}],
\end{eqnarray}
\begin{eqnarray}
&&  \frac{q_{31}-q_{30}}{h}= \nonumber\\
&& \frac{1}{8}[\frac{H(00000010)-H(00000000)}{p_{31}-p_{30}} \nonumber\\
&& +\frac{H(01000010)-H(01000000)}{p_{31}-p_{30}} \nonumber\\
&& +\frac{H(01000110)-H(01000100)}{p_{31}-p_{30}} \nonumber\\
&& +\frac{H(11000110)-H(11000100)}{p_{31}-p_{30}} \nonumber\\
&& +\frac{H(11001110)-H(11001100)}{p_{31}-p_{30}} \nonumber\\
&& +\frac{H(11011110)-H(11011100)}{p_{31}-p_{30}} \nonumber\\
&& +\frac{H(11011111)-H(11011101)}{p_{31}-p_{30}} \nonumber\\
&& +\frac{H(11111111)-H(11111101)}{p_{31}-p_{30}}],
\end{eqnarray}
\begin{eqnarray}
&&\frac{q_{41}-q_{40}}{h}= \nonumber\\
&& \frac{1}{8}[\frac{H(00000001)-H(00000000)}{p_{41}-p_{40}} \nonumber\\
&& +\frac{H(00100001)-H(00100000)}{p_{41}-p_{40}} \nonumber\\
&& +\frac{H(00100011)-H(00100010)}{p_{41}-p_{40}} \nonumber\\
&& +\frac{H(01100011)-H(01100010)}{p_{41}-p_{40}} \nonumber\\
&& +\frac{H(01100111)-H(01100110)}{p_{41}-p_{40}} \nonumber\\
&& +\frac{H(11100111)-H(11100110)}{p_{41}-p_{40}} \nonumber\\
&& +\frac{H(11101111)-H(11101110)}{p_{41}-p_{40}} \nonumber\\
&& +\frac{H(11111111)-H(11111110)}{p_{41}-p_{40}}];
\end{eqnarray}
\begin{eqnarray}
&&\frac{p_{11}-p_{10}}{h}= \nonumber\\
&& -\frac{1}{8}[\frac{H(10000000)-H(00000000)}{q_{11}-q_{10}} \nonumber\\
&& +\frac{H(10001000)-H(00001000)}{q_{11}-q_{10}} \nonumber\\
&& +\frac{H(10011000)-H(00011000)}{q_{11}-q_{10}} \nonumber\\
&& +\frac{H(10011001)-H(00011001)}{q_{11}-q_{10}} \nonumber\\
&& +\frac{H(10111001)-H(00111001)}{q_{11}-q_{10}} \nonumber\\
&& +\frac{H(10111011)-H(00111011)}{q_{11}-q_{10}} \nonumber\\
&& +\frac{H(11111011)-H(01111011)}{q_{11}-q_{10}} \nonumber\\
&& +\frac{H(11111111)-H(01111111)}{q_{11}-q_{10}}],
\end{eqnarray}
\begin{eqnarray}
&&\frac{p_{21}-p_{20}}{h}= \nonumber\\
&& -\frac{1}{8}[\frac{H(01000000)-H(00000000)}{q_{21}-q_{20}} \nonumber\\
&& +\frac{H(01000100)-H(00000100)}{q_{21}-q_{20}} \nonumber\\
&& +\frac{H(11000100)-H(10000100)}{q_{21}-q_{20}} \nonumber\\
&& +\frac{H(11001100)-H(10001100)}{q_{21}-q_{20}} \nonumber\\
&& +\frac{H(11011100)-H(10011100)}{q_{21}-q_{20}} \nonumber\\
&& +\frac{H(11011101)-H(10011101)}{q_{21}-q_{20}} \nonumber\\
&& +\frac{H(11111101)-H(10111101)}{q_{21}-q_{20}} \nonumber\\
&& +\frac{H(11111111)-H(10111111)}{q_{21}-q_{20}}],
\end{eqnarray}
\begin{eqnarray}
&&\frac{p_{31}-p_{30}}{h}= \nonumber\\
&&-\frac{1}{8}[\frac{H(00100000)-H(00000000)}{q_{31}-q_{30}} \nonumber\\
&& +\frac{H(00100010)-H(00000010)}{q_{31}-q_{30}} \nonumber\\
&& +\frac{H(01100010)-H(01000010)}{q_{31}-q_{30}} \nonumber\\
&& +\frac{H(01100110)-H(01000110)}{q_{31}-q_{30}} \nonumber\\
&& +\frac{H(11100110)-H(11000110)}{q_{31}-q_{30}} \nonumber\\
&& +\frac{H(11101110)-H(11001110)}{q_{31}-q_{30}} \nonumber\\
&& +\frac{H(11111110)-H(11011110)}{q_{31}-q_{30}} \nonumber\\
&& +\frac{H(11111111)-H(11011111)}{q_{31}-q_{30}}],
\end{eqnarray}
\begin{eqnarray}\label{14}
&&\frac{p_{41}-p_{40}}{h}= \nonumber\\
&& -\frac{1}{8}[\frac{H(00000010)-H(00000000)}{q_{41}-q_{40}} \nonumber\\
&& +\frac{H(00010001)-H(00000001)}{q_{41}-q_{40}} \nonumber\\
&& +\frac{H(00110001)-H(00100001)}{q_{41}-q_{40}} \nonumber\\
&& +\frac{H(00110011)-H(00100011)}{q_{41}-q_{40}} \nonumber\\
&& +\frac{H(01110011)-H(01100011)}{q_{41}-q_{40}} \nonumber\\
&& +\frac{H(01110111)-H(01100111)}{q_{41}-q_{40}} \nonumber\\
&& +\frac{H(11110111)-H(11100111)}{q_{41}-q_{40}} \nonumber\\
&& +\frac{H(11111111)-H(11101111)}{q_{41}-q_{40}}].
\end{eqnarray}
The discretization of each partial  derivative of the Hamiltonian
with respect to one of the position and momentum variables is the
average of eight Hamiltonian difference terms. It is easy to find
that the left-hand side of $\sum^{4}_{i=1}$ [Equation $(5+i)$
$\cdot$ $(p_{i1}-p_{i0})$ $-$ Equation $(9+i)$ $\cdot$
$(q_{i1}-q_{i0})$] vanishes, and the right-hand side is
$H(11111111)-H(00000000)$--that is, the discrete Equations
(6)-(13) of the Hamiltonian Equations (1) and (2) exactly satisfy
the energy-conserving condition (5). These Hamiltonian
differencing symmetric forms  in the right-hand sides of Equations
(6)-(13) possess a second-order accuracy. There is not a systematic
method by means of which the difference equations 6-13 have been
constructed.

The difference schemes of Equations (6)-(13) for 8-dimensional Hamiltonian
problems are not symplectic although they are energy conservative.
If Equations (1) and (2) are nonlinear, Equations (6)-(13) should
be solved by an iterative method, such as a Newton iterative
scheme. However, the iterative solution does not converge when one
or more denominators of the right-hand sides in Equations.
(6)-(13) tend to zero or are sufficiently small. In order to avoid
the occurrence of numerical singularities, we should rewrite as
much as possible each Hamiltonian difference term so that all same
factors between the denominator and the numerator are eliminated
in the Hamiltonian difference. If some singularities or
sufficiently small denominators still arise, the difference of a
certain function is replaced with the partial derivative of the
function. More details on how to handle this kind of numerical
singularities were provided by Bacchini et al. (2018a).

In what follows, we are interested in evaluating the performance
of the new energy-conserving (EC) scheme applied to two models.
For comparison, an RK method, an implicit midpoint
symplectic (IS) method  (Feng 1986; Zhong et al. 2010; Mei et al.
2013) and an extended phase-space explicit symplectic-like (ES)
integrator  (Pihajoki 2015; Liu et al. 2016; Luo et al. 2017; Li
$\&$ Wu 2017) are employed. The extended phase-space method is
simply described in the Appendix.

\section{FPU $\beta$ lattice}

An FPU system (Skokos et al. 2008; Gerlach et
al. 2012) with $N$ dimensions describes the motion of $N$
particles interaction each other. An FPU $\beta$ lattice is a
perturbed FPU system. For our purpose, we take into account an
8-dimensional FPU $\beta$ system with four degrees of freedom as
follows:
\begin{eqnarray}\label{11}
H(\textbf{q},\textbf{p}) &=&
\sum\limits_{i=1}^4\frac{p_{i}^{2}}{2}
+\sum\limits_{i=0}^4[\frac{(q_{i+1}-q_{i})^{2}}{2} \nonumber
\\ & & +\beta\frac{(q_{i+1}-q_{i})^{4}}{4}].
\end{eqnarray}
Boundary conditions are $q_{0}=q_{5}=0$, and $\beta$ is a
non-negative parameter.

When the new energy-conserving method EC is applied to this
system, its implementation is given here. A key point lies in that
singularities or small denominators in the difference terms in the
right-hand sides of Equations (6)-(13) should be eliminated as
much as possible. For example, the difference term in the
right-hand side of Equation (6) is 
$[H(00001000)-H(00000000)]/(p_{11}-p_{10})=(p_{11}+p_{10})/2$. As
another example,
$[H(10011000)-H(00011000)]/(q_{11}-q_{10})=[1+\beta(q^{2}_{11}+q^{2}_{10})/2]
(q_{11}+q_{10})/2 +(q_{11}+q_{10}-2q_{20})\{1
+\beta[(q_{20}-q_{11})^{2}+(q_{20}-q_{10})^{2}]/2\}/2$ in Equation
(10). In this way, all denominators in the right-hand sides of
Equations (6)-(13) are no longer present. Hence, with the Newton
iterative method the iterative solutions of Equations (6)-(13)
acting on the system of Equation (14) have no difficulty.

Taking $\beta=1.5$ and the time step of $h=0.01$, we choose two
orbits whose initial conditions are $\mathbf{q}=(0.1,0.1,0.2,0.2)$
for orbit 1, and $\mathbf{q}=(0.1,0.1,0.2,1.1)$ for orbit 2. As
shown in Figure 1, the RK method shows a secular growth in the
Hamiltonian error, but the new method EC, the implicit symplectic
method IS and the extended phase-space method ES do not. The
latter three methods make the energy stable. In this sense, they
are regarded as energy-conserving schemes. As to the numerical
accuracy of the energy, the RK method is the worst, the IS and ES
integrators are almost the same, and the new algorithm EC is the
best. In particular, the energy error is smaller in several orders
for EC than for IS or ES. It is clear that the new
energy-conserving method is greatly superior to both the implicit
symplectic method and the extended phase-space method in the
conservation of the Hamiltonian.

Figures 1(a) and (b) show that the energy error of orbit 1 given by the new method EC is two orders less than that of orbit 2.
 This is due to the two orbits having
different dynamical behaviors. The maximum Lyapunov exponents
$\lambda$ in Figure 2 (a) and the fast Lyapunov indicators
$\Lambda$ in Figure 2 (b) show the regularity of orbit 1 and the
chaoticity of orbit 2. Here, the Lyapunov exponents are calculated
in terms of the two-particle method (Wu $\&$ Huang 2003). A
bounded orbit is chaotic if its Lyapunov exponent tends to a
stabilized positive value but is regular when the maximum Lyapunov
exponent is zero. The fast Lyapunov indicators are also based on
the idea of the two-particle method (Wu et al. 2006). If this
indicator grows in a power law with time $\log_{10} t$, the
bounded orbit is ordered. However, it is chaotic if this indicator
grows in an exponential law. According to these criteria for the
Lyapunov exponents and the fast Lyapunov indicators distinguishing
between the two cases of order and chaos, we can easily determine
the dynamical features of orbits 1 and 2 in Figure 2 (a) and (b).

Note that the energies of orbits 1 and 2 are 0.031
and 1.815, respectively. This shows that chaos occurs easily for a
large energy. Using the newly proposed algorithm through many
numerical tests, we find that under the present circumstances,
chaos is absent for the energy smaller than 0.5, whereas it is present
for the energy larger than 0.5. Table 1 and Figure 3 relate to the
regularity and chaoticity of some orbits.

\section{Post-Newtonian Hamiltonian of compact binaries with One body spinning}

Various orbits in a post-Newtonian Hamiltonian of compact binaries
with one body spinning are used to test the numerical performance
of the new algorithm. These orbits include circular, spherical,
quasi-spherical, and eccentric orbits.

\subsection{Dynamical model}

Since the novel algorithm is only limited to the use of
eight-dimensional systems, we consider a compact binary system in
which only one body spins. This system consists of two black holes
with masses $m_{1}$ and $m_{2}$. The total mass is
$M=m_{1}+m_{2}$. We take a reduce mass of $\mu=m_{1}m_{2}/M$, a mass
ratio of $\gamma=m_{1}/m_{2}$ and a mass parameter of 
$\eta=\mu/M=\gamma/(1+\gamma)^{2}$. $\textbf{r}=(x,y,z)$ is a
coordinate of body 1 relative to body 2, and
$\textbf{n}=\textbf{r}/r$ is a radial unit vector, where
$r=|\mathbf{r}|$ is a radius. We suppose that the spin motion of body 1 is
described by $\textbf{S}_{1}$. The speed of light, $c$, and the
constant of gravity, $G$, are one geometric unit, $c=G=1$. The
dynamics of compact binaries can be described by the following
post-Newtonian (pn) Hamiltonian (Damour et al. 2000a, 2000b, 2001;
Nagar 2011):
\begin{eqnarray}\label{9}
 H(\textbf{r},\textbf{p},\textbf{S}_{1}) &=& H_{o}(\textbf{r},\textbf{p})
 + H_{so}(\textbf{r},\textbf{p},\textbf{S}_{1})\nonumber \\
 && + H_{ss}(\textbf{r},\textbf{S}_{1}).
\end{eqnarray}
$H_{o}$ is an orbital part, consisting of the Newtonian,
the first-order post-Newtonian (1pn), and the second-order post-Newtonian
(2pn) terms:
\begin{equation}
H_{o} =H_{n}(\textbf{r},\textbf{p})+H_{1pn}(\textbf{r},\textbf{p})+H_{2pn}(\textbf{r},\textbf{p}),
\end{equation}
\begin{equation}
H_{n} = \frac{\textbf{p}^2}{2}-\frac{1}{r},
\end{equation}
\begin{eqnarray}
H_{1pn} &=& \frac{1}{8}(3\eta-1)\textbf{p}^4-\frac{1}{2r}[(3+\eta)\textbf{p}^2 \nonumber \\
&& +\eta(\textbf{n}{\cdot}\textbf{p})^2]+\frac{1}{2r^2},
\end{eqnarray}
\begin{eqnarray}
 H_{2pn} &=& \frac{1}{16}(1-5\eta+5\eta{^2})\textbf{p}^6 +\frac{1}{8r}[(5-20\eta \nonumber \\
&& -3\eta^2)\textbf{p}^4 -2\eta^2(\textbf{n}\cdot\textbf{p})^2\textbf{p}^2 \nonumber \\
&& -3\eta^2(\textbf{n}\cdot\textbf{p})^4] +\frac{1}{2r^2}[(5+8\eta)\textbf{p}^2  \nonumber \\
&& +3\eta(\textbf{n}\cdot\textbf{p})^2]-\frac{1}{4r^3}(1+3\eta).
\end{eqnarray}
Spin effects have spin-orbit couplings of
$H_{so}(\textbf{r},\textbf{p},\textbf{S}_{1})$, and spin-spin
couplings of $H_{ss}(\textbf{r},\textbf{S}_{1})$. The two types of
spin contributions are written as
\begin{eqnarray}
H_{so} &=& \frac{\eta}{r^{3}}(\textrm{g}_{\alpha}+\frac{1}{\gamma}
\textrm{g}_{\beta})\textbf{S}_{1}\cdot \textbf{L},   \\
H_{ss} &=& \frac{\eta}{2r^{3}}[3(\textbf{S}_{\gamma}\cdot
\textbf{n})^{2}-\textbf{S}_{\gamma}^{2}],
\end{eqnarray}
\begin{eqnarray}
\textrm{g}_{\alpha} &=&
2+[\frac{19}{8}\eta\textbf{p}^{2}+\frac{3}{2}\eta
(\textbf{n}\cdot \textbf{p})^{2} \nonumber \\
&& -(6+2\eta)\frac{1}{r}], \\
\textrm{g}_{\beta} &=& \frac{3}{2} -(\frac{5}{8}+2\eta)\textbf{p}^{2}
+\frac{3}{4}\eta (\textbf{n}\cdot \textbf{p})^{2} \nonumber \\
&& -\frac{1}{r}(5+2\eta), \\
\textbf{S}_{\gamma} &=& (1+\frac{1}{\gamma})\textbf{S}_{1}, \\
\textbf{L} &=& \textbf{r}\times \textbf{p}.
\end{eqnarray}
Here, the spin-orbit couplings have 1.5pn and 2.5pn terms and the
spin-spin couplings are at 2pn order. The Newton-Wigner-Pryce spin
supplementary condition $\kappa = 0$ (Mik$\acute{o}$czi 2017) is
used. In addition, the Hamiltonian (15) uses dimensionless
operations: $\mathbf{r}\rightarrow M \mathbf{r}$, $t\rightarrow M
t$, $H\rightarrow \mu H$, $\mathbf{L}\rightarrow M \mu
\mathbf{L}$, and $\mathbf{S}_1\rightarrow M \mu \mathbf{S}_1$,
i.e.,
\begin{equation}
  \textbf{S}_{1}=S_{1}\hat{\textbf{S}}_{1}, ~~~~
S_{1}=\frac{\chi_{1}m_{1}^{2}}{\mu M}.
\end{equation}
$\chi_1$ is a dimensionless parameter given at the interval [0,
1], and $\hat{\textbf{S}}_{1}$ is a unit spin vector.

The position and momentum variables, $\mathbf{r}$ and
$\mathbf{p}$, are conjugate because they satisfy the Hamiltonian
canonical equations (1) and (2). However, the spin variable, 
$\textbf{S}_{1}$, is not because its evolution equation is a
non-canonical equation
\begin{equation}
  \frac{d\textbf{S}_{1}}{dt}=\frac{\partial H}{\partial
  \textbf{S}_{1}} \times \textbf{S}_{1}.
\end{equation}
In this case, the application of the new algorithm of Equations (6)-(13) to the
system (15) becomes difficult. This problem can be solved with the
aid of the canonical, conjugate spin cylindrical-like coordinates
$\theta$ and $\xi$ constructed by Wu $\&$ Xie (2010):
\begin{equation}
\textbf{S}_{1} =(\rho\cos\theta,\rho\sin\theta,\xi)^{\intercal},
~~~~ \rho =\sqrt{S^{2}_{1}-\xi^{2}}.
\end{equation}
It is clear that $\textbf{S}_{1}$ is a two-dimensional vector with
three components. The Hamiltonian (15) is completely canonical and
has 8-dimensional phase-space variables, 
$\mathbf{q}^{*}=(\mathbf{q},\theta)$ and
$\mathbf{p}^{*}=(\mathbf{p},\xi)$. This canonical system is
labeled as
\begin{equation}
\mathcal{H}(\mathbf{q}^{*},\mathbf{p}^{*})=
H(\mathbf{q},\mathbf{p},\mathbf{S}_1),
\end{equation}
which has four independent integrals: the total energy (i.e.
$\mathcal{H}$)  and three components of the total angular momentum
vector, $\mathbf{J}=\mathbf{L}+\mathbf{S}_1$. Therefore,
$\mathcal{H}$ is integrable and non-chaotic.

\subsection{Numerical investigations}

Without doubt, the new energy-conserving algorithm given by
Equations (6)-(13) is suitable for the canonical system
$\mathcal{H}$ in Equation (29). Problems of numerical
singularities or small denominators may frequently occur but can
be solved in terms of the suggestions given in Section 2.

Taking the parameters $\chi_1=1$, $\gamma=1$ and initial
conditions $y=z=p_x=p_z=\theta=\xi=0$ and $p_y=\sqrt{(1-e)/x}$, we
plot Figure 4(a) in which the Hamiltonian errors of the above
algorithms solving orbit 1 in Table 2 are shown. The new method
EC, the implicit symplectic integrator IS, and the extended
phase-space method ES can make the Hamiltonian errors stable. The
energy error for EC is of the order of $~$ $10^{-15}$, and those for
IS and ES remain of the order of $10^{-6}$. The energy error for RK
increases with time and reaches about an order of $10^{-5}$ when
the integration time arrives at $10^{6}$. These results are what
we expect and seem to be independent of the initial eccentricity,
$e$. See Table 2 for more information. The  results are obtained
from large initial orbital radii and different time steps. For a
small initial radius and a fixed time step, e.g. $r=x=40$ and
$h=0.1$ in Table 3, these algorithms that act on some orbits show
a similar performance. Of course, there are some differences. The
energy errors for the three methods, RK, IS, and ES, are closely
associated to the initial eccentricity and become large with the
increasing eccentricity, but the energy error for the new scheme
EC is not very sensitive dependent on the initial eccentricity and
still remains of the order of $10^{-14}$ after the integration time
$10^{5}$. These results in Tables 2 and 3 are given when the
spin-orbit couplings and the spin-spin couplings are included.
What about relative position errors of these algorithms? When more
accurate solutions given by an eighth- and ninth-order
Runge-Kutta-Fehlberg algorithm of variable step sizes are taken as
reference solutions, the relative position errors that these
methods show can be obtained in Figure 4(b). As the integration
times are long enough and a larger time step is adopted, the
position errors remain stable for the new method EC, whereas they grow
with time for the other three methods. Particularly for the RK
method, its  position errors are larger than for EC after a long
enough time.

It is not only at the starting time but also at any time that the
eccentricity of orbit 10 in Table 3 is always identical to zero in
Equation (16) when no body spins. In fact, this orbit is an exact
circular orbit. This is because it satisfies the conditions of
circular orbits on the equatorial plane: $dr/dt=\partial
H_o/\partial p_r=0$, $p_r=0$, and $\partial H_o/\partial r=0$,
where $p_r=\textbf{n}\cdot \textbf{p}$ denotes a radial momentum.
In this case,
$\textbf{p}^2=p^{2}_{r}+L^{2}_{z}/r^{2}=L^{2}_{z}/r^{2}$, where
$L_{z}=xp_y=6.6473$ is the $z$ component of the orbital angular
momentum, $\textbf{L}$. The angular frequency of the circular orbit
is $\omega_o=\partial H_o/\partial L_{z}=0.0038$. Such a circular
orbit is used to check the numerical performance of these methods.
Because this orbit is only limited to staying at a six-dimensional
phase space of the system in equation (16), the energy-conserving
method of Bacchini et al. (2018a) rather than the new
energy-conserving scheme is suitable for integrating this orbit.
The four methods give the energy errors (not plotted) to the
circular orbit, like those to orbit 1 in Figure  4(a). They almost
remain of the radius $r=40$ of the circular orbit in Figure  5(a)
when the integration times are short. However, the position errors
for RK will be larger than for the EC method as the integration
times last long enough in Figure  5(b). This result looks like
that of orbit 1 in Figure  4(b). The EC method does not show a
secular growth in the position errors in Figures 4(b) and 5(b). Of
course, there is a typical difference between Figures 4(b) and
5(b) that the implicit symplectic integrator IS and the extended
phase-space method ES have a secular growth in the position errors
of orbit 1 but do not have in the position errors of the circular
orbit. When the spin-orbit couplings in Equation (20) are added to
the orbital part (16), the term $H_{so}$ is a conserved quantity
and does not contain $p_r$. Therefore, $dr/dt=0$ is still
existent. The radius $r=40$ remains invariant, but the orbit is a
spherical orbit rather than a circular orbit because of the spin
of body 1 leading to the precession of orbits. This spherical
orbit with the angular frequency $\omega_{so}=\omega_o+\partial
H_{so}/\partial L_{z}=0.0038$ is an orbit in a eight-dimensional
phase space and so our new EC method rather than  the
energy-conserving method of Bacchini et al. (2018a) becomes
useful. The four algorithms show the preference of such a
spherical orbit during a short integration time in Figure 5(c).
The position errors in Figure 5(d) are also similar to those in
Figure 5(b). A difference lies in that the position errors are
smaller for RK than for EC. When the spin-spin interactions in
Equation (21) are also included, these methods almost give same
Figure 5(e) that the radius $r$ oscillates around 40.025 in a
small amplitude and the spherical orbit is slightly destroyed.
This destruction of the spherical orbit is not large because the
spin-spin effects are small compared to the spin-orbit ones. With
the inclusion of the spin-spin couplings, the spherical orbit
yielded by  the spin-orbit couplings becomes a quasi-spherical
orbit. The position errors given by these schemes for the
quasi-spherical orbit  in Figure 5(f) are almost the same as those
for the spherical orbit in Figure 5(d). If the time step $h=1$ in
Figure 5 is replaced with a small time step $h=0.01$ in Table 4,
the relative position errors are the smallest for ES but the
largest for EC. RK and IS have almost the same errors.

Now, let eccentrical orbits (e.g.  the eccentrical orbit 11 in
Table 3) be used as tested orbits. For this case, the energy
errors (not plotted) that the four algorithms show are nearly the
same as those in Figure 4(a). The relative position errors have
secular growths for the four algorithms, as shown in Figure 6.
They are larger for RK than for EC after a long enough time.

Seen from the above numerical experiments, the new method is the
most effective to conserve energies compared to the other three
schemes. However, it is not superior to the RK method in
the accuracy of numerical solutions when the time step is small
and the integration time is not long enough, as shown in Figures
4(b), 5(b), (d), (f), and 6 and Table 4. In other words, such
energy-conserving integrators are typically characterized by
larger trajectory errors. This is because numerical errors that
are prevented in the energy are then reversed to the position. For
periodic or bounded motion, such errors can be typically ignored,
since they will mostly result in frequency/phase mismatches
without causing the disruption of the bounded orbit. On the other
hand, the energy-conserving integrators are nonsymplectic, hence
they are not characterized by conservation of phase-space
trajectories. The highly geometric character of these integrators
suffices in preserving phase space trajectories to a high degree.
It is, however, not impossible that some geometric features of the
trajectory may suffer from additional numerical errors, which
would be absent in symplectic integrators. Besides this drawback,
the energy-conserving integrators need much additional
computational cost, since they are implicit and involve a system
of nonlinear equations that are handled during the computation. This
fact is conformed in Table 5.

\section{Summary}

The novel energy-conserving method given by Equations (6)-(13) is
specifically designed for an eight-dimensional Hamiltonian system
with four degrees of freedom. In this algorithm, eight partial
derivatives of the Hamiltonian with respect to each phase-space
variable are discretized and the discretization of the partial
derivatives is the average of eight Hamiltonian difference terms.
This average gives a second-order accuracy to the Hamiltonian
derivative. This algorithm is implicit and can be solved with the
aid of the Newton iterative method when the Hamiltonian is
nonlinear. It is exactly energy conserving from the theoretical
viewpoint but is nonsymplectic.

When the FPU-$\beta$ lattice is chosen as a tested model, the
newly proposed method is shown to have extremely good numerical
performance in the conservation of energy. Regardless of whether
the considered orbit is regular or chaotic, this new algorithm is
greatly superior to the RK method, the implicit midpoint
symplectic method, and the extended phase-space explicit
symplectic-like integrator. Of course, the energy-conserving
accuracy of the new algorithm is better for the ordered case than
for the chaotic case. Using the new energy-conserving method, we
find several regular orbits and chaotic ones in the system. The
threshold of energies between order and chaos can also be found
under a certain circumstance. When the post-Newtonian conservative
system of compact binaries with one body spinning is used as
another tested model, the newly proposed method is still extremely
good in the preservation of energies regardless of initial orbital
eccentricities. It is advantageous over very long times and for
large time steps compared to the state-of-the-art RK method
in the accuracy of numerical solutions. It exhibits a higher
computational cost than any one of the other three algorithms.

For some specific purposes on the preservation of
energies in numerical simulations with much longer
times,\footnote{The purposes, e.g., are to obtain higher
accuracies of the semimajor axis, mean motion, and mean anomaly in
the Keplerian problem in the solar system, or to accurately grasp
the frequency of a gravitational wave emitted from the circular,
spherical, or quasi-spherical orbit in relativistic post-Newtonian
systems of compact binaries.} such energy-conserving integrators
are worth recommending for application. The new integrator could be
used to simulate relativistic charged particles moving in a
time-varying external electromagnetic field (P\'{e}tri 2017). Such
a field can be described by an  associated time-dependent
Hamiltonian. This Hamiltonian $H$ has eight dimensions, including three
spatial coordinates + time, three velocity components, and time change
with respect to proper time (i.e. the Lorentz factor). By extending
the phase space of the Hamiltonian, we will obtain a zero
Hamiltonian $\widetilde{H}=H+p_0$, where $p_0$ is a momentum with
respect to time. This would provide a mean to simulate charged
particles in time-varying external fields with exact
zero-Hamiltonian conservation, which is a very desirable feature in the
context of test particle simulations in high-energy astrophysical
scenarios. The new integrator will also be suitable for modeling
a time-dependent spacetime (Bohn et al. 2015). Although the
Hamiltonian $H=\frac{1}{2}g^{\alpha\beta} p_{\alpha}p_{\beta}$ for
the spacetime, $dS^2=g_{\alpha\beta}dx^{\alpha}dx^{\beta}$ does
explicitly depends on the coordinate time, $t$, it has an invariant
quantity, $H=-1/2$. From the theory, this invariant quantity can
still be preserved by the new integrator. With the aid of the
integrator, the characteristic of gravitational waves for
circular, spherical, quasi-spherical, and eccentric orbits in
relativistic post-Newtonian systems of compact binaries will be
analyzed in the future work.

\section*{Acknowledgments}

The authors would like to express their deep gratitude to the
referee for valuable comments and suggestions. This research has
been supported by the National Natural Science Foundation of China
(grant Nos. 11533004, 11973020, 11663005, 11533003, and 11851304),
the Graduate Innovation Foundation of Jiangxi Province (grant No.
YC2018-S002), the Special Funding for Guangxi Distinguished
Professors (2017AD22006), and the Natural Science Foundation of
Guangxi (grant No. 2018GXNSFGA281007).

\section*{APPENDIX}

\subsection*{Extended phase-space explicit leapfrog integrators for inseparable
Hamiltonian problems}

Usually, an explicit second-order leapfrog integrator of Wisdom
$\&$ Holman (1991) is not suitable  for Hamiltonian problems with
inseparable forms of coordinates and momenta. However, it is still
valid if the extended phase-space method of Pihajoki (2015) is
considered.

Although a Hamiltonian $H(\textbf{q},\textbf{p})$ is inseparable,
its modified form is
\begin{equation}
\Gamma(\textbf{q},\tilde{\textbf{q}},\textbf{p},\tilde{\textbf{p}})
=H_{1}(\textbf{q}, \tilde{\textbf{p}}) + H_{2}(\tilde{\textbf{q}},
\textbf{p}),
\end{equation}
where $H_{1}=H_{2}=H$, can be split into two solvable parts,
$H_{1}$ and $H_{2}$. $A(h)$ is an  operator of $H_{1}$, and
$B(h)$ is another  operator of $H_{2}$. The leapfrog algorithm for
the new Hamiltonian $\Gamma$ is
\begin{equation}\label{8}
S(h) = A(\frac{h}{2})B(h)A(\frac{h}{2}).
\end{equation}

Under the same initial conditions, the original solution
$(\textbf{q},\textbf{p})$ and the extended solution
$(\tilde{\textbf{q}},\tilde{\textbf{p}})$ should be the same.
However, their coupled derivatives lead to both solutions having
some differences. To make the two solutions equal, the leapfrog
$S(h)$ needs maps as feedback after the two solutions, e.g.,
\begin{equation}
ES(h)= M\otimes S(h),
\end{equation}
where  $M$ is a map. These permuted maps can be given in various
forms, such as the maps of Pihajoki (2015) and Liu et al. (2016).
As a good choice of the map $M$, the midpoint permutations between
the old variables $(\textbf{q},\textbf{p})$ and the new variables
$(\tilde{\textbf{q}},\tilde{\textbf{p}})$ are given by Luo et al.
(2017):
\begin{eqnarray}
 &&   \frac{\textbf{q}+\tilde{\textbf{q}}}{2} \rightarrow \textbf{q}, \nonumber \\
 && \frac{\textbf{q}+\tilde{\textbf{q}}}{2} \rightarrow
  \tilde{\textbf{q}},  \nonumber \\
 && \frac{\textbf{p}+\tilde{\textbf{p}}}{2} \rightarrow \textbf{p}, \nonumber \\
 && \frac{\textbf{p}+\tilde{\textbf{p}}}{2} \rightarrow   \tilde{\textbf{p}}.
\end{eqnarray}

This algorithm $ES$ is an extended phase-space  explicit leapfrog
integrator. The inclusion of the permuted map makes this algorithm
nonsymplectic. However, this integrator is a symmetric method and
shows no secular growth in the errors of energy.

\newpage

\begin{table}
\begin{center}
\small \caption{Dynamical Features of Several Orbits in the
FPU-$\beta$ System. These orbits have the same initial momentum
$\mathbf{p}=(0,0,0,0)$ but different positions
$\mathbf{q}=(q_1,q_2,q_3,q_4)$. $E=H$ corresponds to the energy of each orbit.} \label{t1}
\begin{tabular}{cccc}\hline
Orbit & Initial Position & $E$ & Dynamics \\
\hline
1 & (0.1,0.1,0.2,0.2) & $0.031$ & order \\
\hline
2 & (0.1,0.1,0.2,1.1) & $1.815$ & chaos \\
\hline
3 & (0.5,0.5,0.5,0.5) & $0.297$ & order \\
\hline
4 & (0.55,0.5,0.5,0.5) & $0.335$ & order \\
\hline
5 & (0.6,0.5,0.5,0.5) & $0.382$ & order \\
\hline
6 & (0.62,0.5,0.5,0.5) & $0.403$ & order \\
\hline
7 & (0.7,0.5,0.5,0.5) & $0.504$ & chaos \\
\hline
8 & (0.75,0.5,0.5,0.5) & $0.581$ & chaos \\
\hline
9 & (0.8,0.5,0.5,0.5) & $0.670$ & chaos \\
\hline
10 & (0.85,0.5,0.5,0.5) & $0.772$ & chaos \\
\hline
\end{tabular}
\end{center}
\end{table}

\begin{table*}
\begin{center}
\caption{After Integration Time $t=10^6$, Hamiltonian Errors
$\Delta H$ of the four algorithms EC, RK, ES, and IS Act on
Several Orbits in the post-Newtonian problem of compact binaries
with One body spinning, Given by Equation (29). The spin-orbit
couplings and spin-spin effects are included together. Initial
values of $x$, $e$ are given below, $py=\sqrt{(1-e)/x}$, and other
initial values are 0. The time step $h$ for each orbit may be
different.} \label{t2}
\begin{tabular}{ccccccccc}\hline
Orbit & $x$  & $e$ & $h$  & EC & RK & ES & IS \\
\hline
1 & 120  & 0.0 & 15 & $8.37 \times 10^{-16}$ & $2.83 \times 10^{-6}$ & $1.53 \times 10^{-8}$ & $3.05 \times 10^{-8}$ \\
\hline
2 & 120  & 0.1 & 5 & $6.79 \times 10^{-16}$ & $2.47 \times 10^{-7}$ & $3.23 \times 10^{-9}$ & $6.46 \times 10^{-9}$ \\
\hline
3 & 120  & 0.2 & 10 & $3.37 \times 10^{-16}$ & $5.77 \times 10^{-6}$ & $1.65 \times 10^{-8}$ & $3.32 \times 10^{-8}$ \\
\hline
4 & 150  & 0.35 & 4 & $1.41 \times 10^{-16}$ & $4.87 \times 10^{-7}$ & $3.10 \times 10^{-9}$ & $6.22 \times 10^{-9}$ \\
\hline
5 & 150  & 0.55 & 5 & $9.29 \times 10^{-16}$ & $2.06 \times 10^{-5}$ & $2.78 \times 10^{-8}$ & $5.40 \times 10^{-8}$ \\
\hline
6 & 180  & 0.6 & 5 & $2.68 \times 10^{-15}$ & $1.33 \times 10^{-5}$ & $2.64 \times 10^{-9}$ & $5.66 \times 10^{-9}$ \\
\hline
7 & 200  & 0.65 & 8 & $1.39 \times 10^{-15}$ & $6.71 \times 10^{-5}$ & $1.48 \times 10^{-6}$ & $2.99 \times 10^{-6}$ \\
\hline
8 & 200  & 0.7 & 2 & $2.88 \times 10^{-15}$ & $3.31 \times 10^{-6}$ & $3.51 \times 10^{-11}$ & $1.20 \times 10^{-10}$ \\
\hline
9 & 240  & 0.8 & 2.5 & $5.74 \times 10^{-15}$ & $2.84 \times 10^{-5}$ & $9.36 \times 10^{-10}$ & $1.71 \times 10^{-11}$ \\
\hline
\end{tabular}
\end{center}
\end{table*}

\begin{table*}
\begin{center}
\small \caption{After Integration Time $t=10^5$, Hamiltonian
Errors $\Delta H$ of the four algorithms EC, RK, ES, and IS Solve 
some orbits in the post-Newtonian problem of compact binaries with
One body spinning, Given by Equation (29). The spin-orbit
couplings and spin-spin effects are included together. Initial
value of $x$ is 40, those of $p_y$ and $e$ are given below, and
other initial values including the initial canonical spin
cylindrical-like coordinates $\theta$ and $\xi$  are 0. The time
step $h=0.1$ is fixed.} \label{t3}
\begin{tabular}{cccccccc}\hline
orbit  & $p_y$ & $e$ & EC & RK & ES & IS \\
\hline
10  & 0.166 & 0.0 & $6.03 \times 10^{-16}$ & $1.34 \times 10^{-10}$ & $2.88 \times 10^{-14}$ & $5.71 \times 10^{-14}$ \\
\hline
11  & 0.158 & 0.0 & $5.07 \times 10^{-15}$ & $5.27 \times 10^{-10}$ & $2.08 \times 10^{-10}$ & $4.16 \times 10^{-10}$ \\
\hline
12  & 0.141 & 0.2 & $2.07 \times 10^{-15}$ & $2.31 \times 10^{-9}$ & $1.27 \times 10^{-9}$ & $2.55 \times 10^{-9}$ \\
\hline
13  & 0.122 & 0.4 & $7.74 \times 10^{-15}$ & $4.60 \times 10^{-8}$ & $2.07 \times 10^{-10}$ & $4.15 \times 10^{-10}$ \\
\hline
14  & 0.100 & 0.6 & $6.88 \times 10^{-15}$ & $1.43 \times 10^{-6}$ & $2.73 \times 10^{-8}$ & $5.49 \times 10^{-8}$ \\
\hline
15  & 0.071 & 0.8 & $1.50 \times 10^{-14}$ & $1.19 \times 10^{-5}$ & $7.16 \times 10^{-10}$ & $1.15 \times 10^{-10}$ \\
\hline
\end{tabular}
\end{center}
\end{table*}

\begin{table*}
\begin{center}
\small \caption{Relative Position Errors for the Four Algorithms that Solve the Circular, Spherical, and Quasi-spherical orbits in
Figure 5. Here, the time step $h=0.01$ is unlike the time step
$h=1$ in Figure 5 (b), (d) and (f).} \label{t4}
\begin{tabular}{ccccc}\hline
 &  & Circular Orbit &  & \\
\hline
Steps & EC & RK & ES & IS \\
\hline
300 & $6.9 \times 10^{-10}$ & $7.0 \times 10^{-14}$ & $2.0 \times 10^{-14}$ & $3.7 \times 10^{-14}$ \\
\hline
600 & $1.4 \times 10^{-10}$ & $2.8 \times 10^{-13}$ & $7.2 \times 10^{-14}$ & $1.4 \times 10^{-13}$ \\
\hline
900 & $2.1 \times 10^{-9}$ & $6.4 \times 10^{-13}$ & $1.5 \times 10^{-13}$ & $3.2 \times 10^{-13}$ \\
\hline
1800 & $4.6 \times 10^{-9}$ & $2.5 \times 10^{-12}$ & $6.4 \times 10^{-13}$ & $1.3 \times 10^{-12}$ \\
\hline
2000 & $5.2 \times 10^{-9}$ & $3.1 \times 10^{-12}$ & $7.9 \times 10^{-13}$ & $1.6 \times 10^{-12}$ \\
\hline
 &  & Spherical Orbit &  & \\
\hline
300 & $3.4 \times 10^{-10}$ & $7.0 \times 10^{-14}$ & $2.0 \times 10^{-14}$ & $3.7 \times 10^{-14}$ \\
\hline
600 & $6.0 \times 10^{-10}$ & $2.8 \times 10^{-13}$ & $7.2 \times 10^{-14}$ & $1.4 \times 10^{-14}$ \\
\hline
900 & $7.8 \times 10^{-10}$  & $6.4 \times 10^{-13}$ & $1.6 \times 10^{-13}$ & $3.2 \times 10^{-13}$ \\
\hline
1800 & $6.0 \times 10^{-13}$ & $2.5 \times 10^{-12}$ & $6.4 \times 10^{-13}$ & $1.3 \times 10^{-12}$ \\
\hline
2000 & $5.4 \times 10^{-10}$ & $3.1 \times 10^{-12}$ & $7.9 \times 10^{-13}$ & $1.6 \times 10^{-12}$ \\
\hline
 &  & Quasi-Spherical orbit &  & \\
\hline
300 & $3.4 \times 10^{-10}$ & $7.0 \times 10^{-14}$ & $1.7 \times 10^{-14}$ & $3.5 \times 10^{-14}$ \\
\hline
600 & $6.1 \times 10^{-10}$ & $2.8 \times 10^{-13}$ & $7.5 \times 10^{-14}$ & $1.4 \times 10^{-13}$ \\
\hline
900 & $7.8 \times 10^{-10}$ & $6.4 \times 10^{-13}$ & $1.7 \times 10^{-13}$ & $3.2 \times 10^{-13}$ \\
\hline
1800 & $7.9 \times 10^{-12}$ & $2.5 \times 10^{-12}$ & $6.4 \times 10^{-13}$ & $1.3 \times 10^{-12}$ \\
\hline
2000 & $5.3 \times 10^{-10}$ & $3.1 \times 10^{-12}$ & $7.9 \times 10^{-13}$ & $1.6 \times 10^{-12}$ \\
\hline
\end{tabular}
\end{center}
\end{table*}

\begin{table}
\begin{center}
\small \caption{CPU times (unit: Seconds) for the Four Methods
Solving Various Orbits with Different Time Steps $h$. The
integration times are $t=10^6$ for orbits 1-9 and $t=10^5$ for
orbits 10-15.} \label{t5}
\begin{tabular}{cccccc}\hline
Orbit & $h$ & EC & RK & ES & IS  \\
\hline
1 & $15$ & $59$ & $1$ & $1$ & $2$ \\
\hline
2 & $5$ & $176$ & $3$ & $4$ & $6$ \\
\hline
3 & $10$ & $86$ & $1$ & $2$ & $3$ \\
\hline
4 & $4$ & $214$ & $3$ & $5$ & $7$ \\
\hline
5 & $5$ & $171$ & $3$ & $4$ & $6$ \\
\hline
6 & $5$ & $172$ & $3$ & $4$ & $6$ \\
\hline
7 & $8$ & $107$ & $2$ & $2$ & $4$ \\
\hline
8 & $2$ & $428$ & $6$ & $9$ & $13$ \\
\hline
9 & $2.5$ & $342$ & $5$ & $8$ & $11$ \\
\hline
10 & $0.1$ & $905$ & $13$ & $19$ & $25$ \\
\hline
11 & $0.1$ & $918$ & $13$ & $19$ & $25$ \\
\hline
12 & $0.1$ & $890$ & $13$ & $19$ & $25$ \\
\hline
13 & $0.1$ & $866$ & $13$ & $19$ & $26$ \\
\hline
14 & $0.1$ & $841$ & $13$ & $19$ & $27$ \\
\hline
15 & $0.1$ & $848$ & $13$ & $19$ & $27$ \\
\hline
\end{tabular}
\end{center}
\end{table}

\newpage

\begin{figure*}
\center{
\includegraphics[scale=0.4]{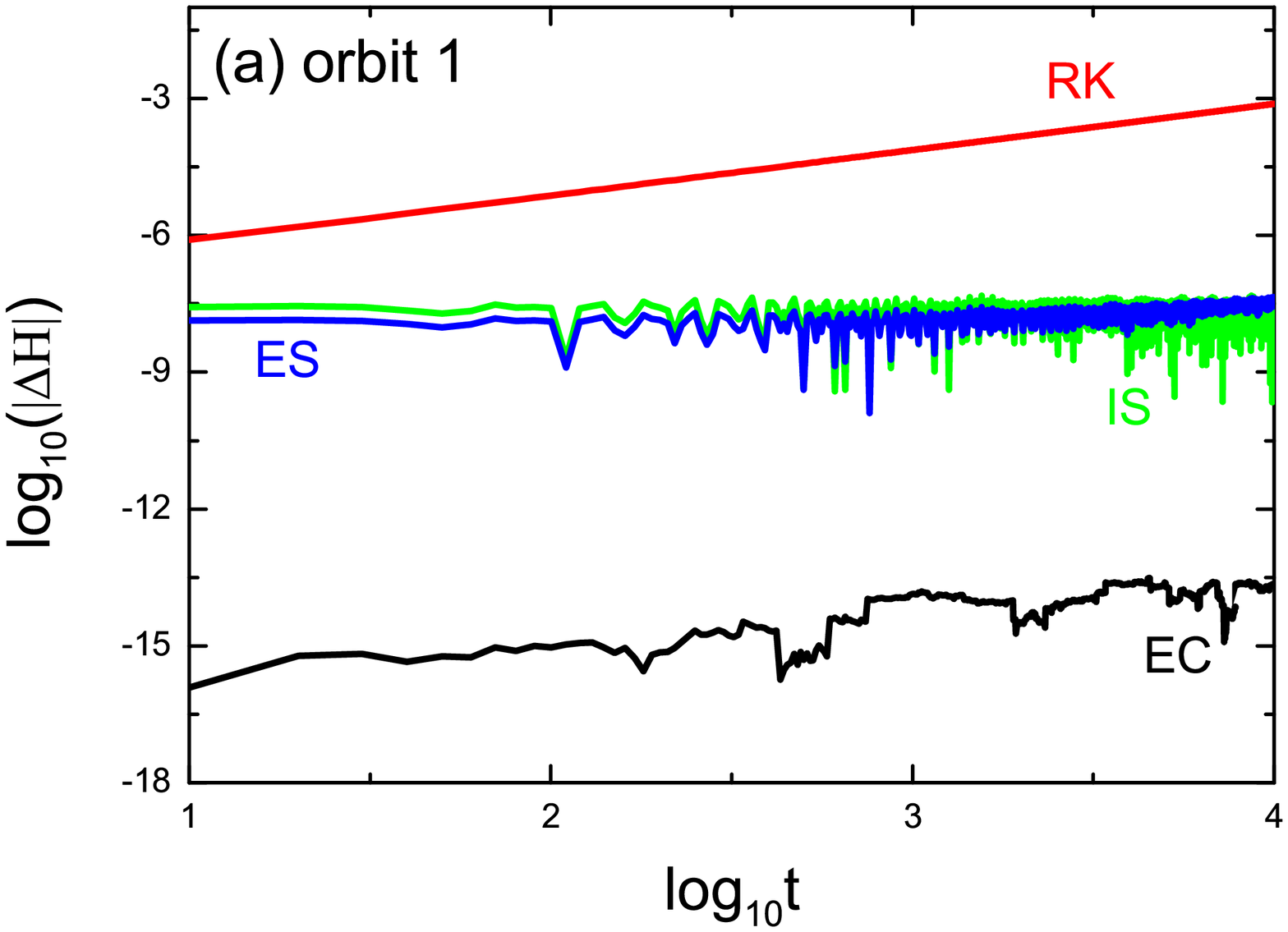}
\includegraphics[scale=0.4]{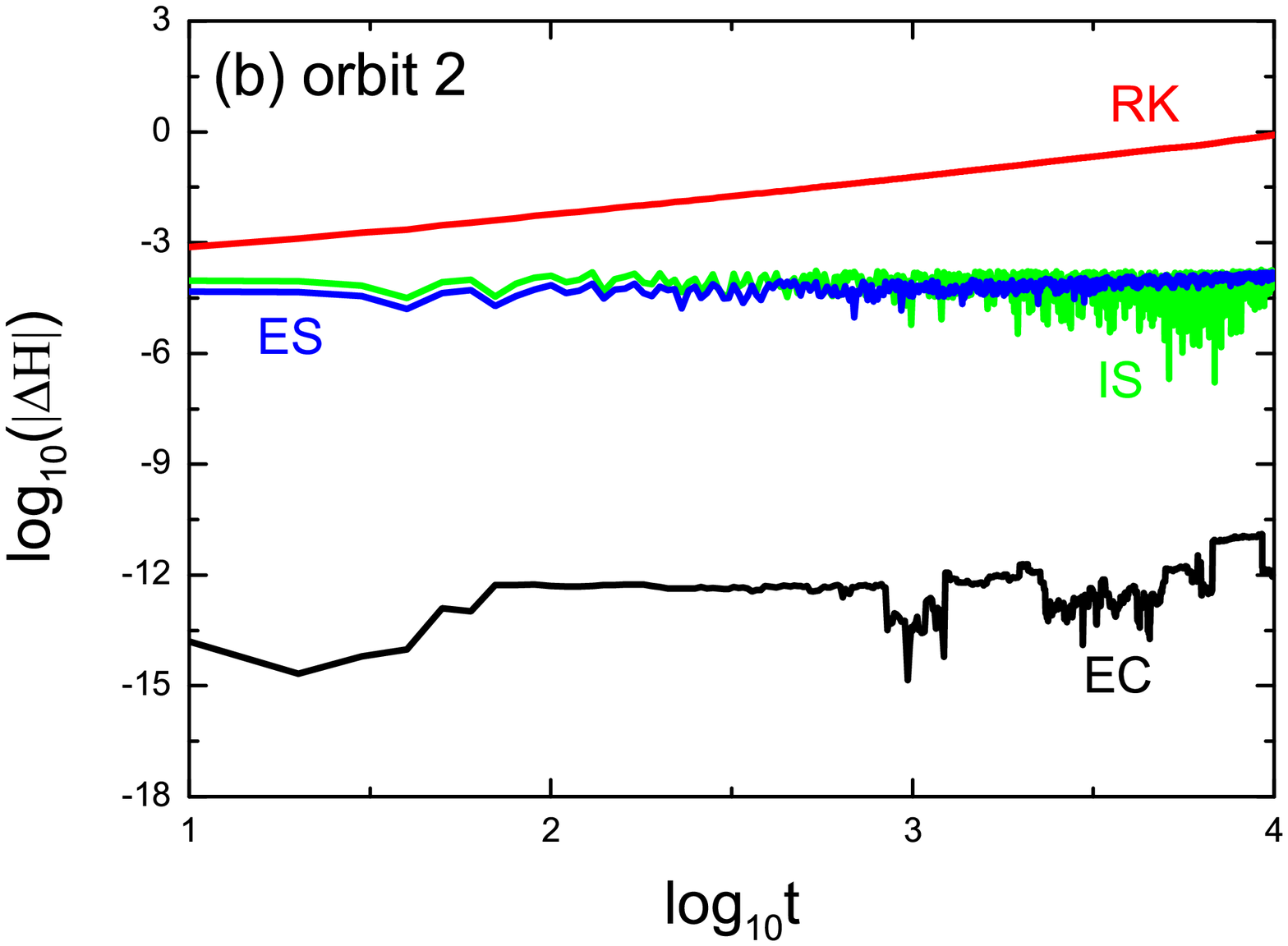}
\caption{Hamiltonian errors for several integrators solving orbits
1 and 2 in the FPU-$\beta$ system. EC, RK, ES, and IS correspond to
the new energy-conserving scheme, the RK method, the
implicit midpoint symplectic algorithm, and the extended
phase-space explicit symplectic-like integrator, respectively.}}
\label{fig1}
\end{figure*}

\begin{figure*}
\center{
\includegraphics[scale=0.4]{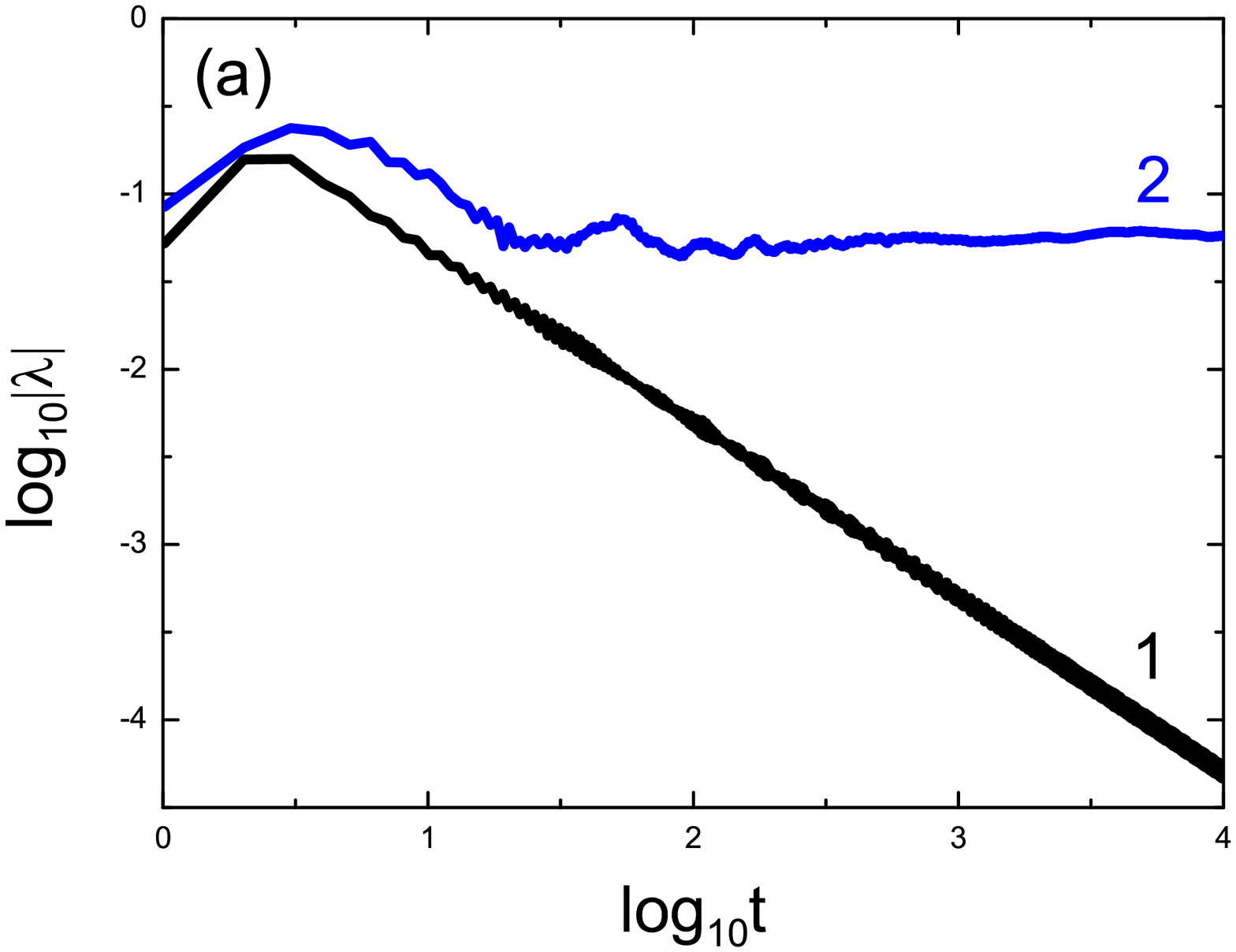}
\includegraphics[scale=0.4]{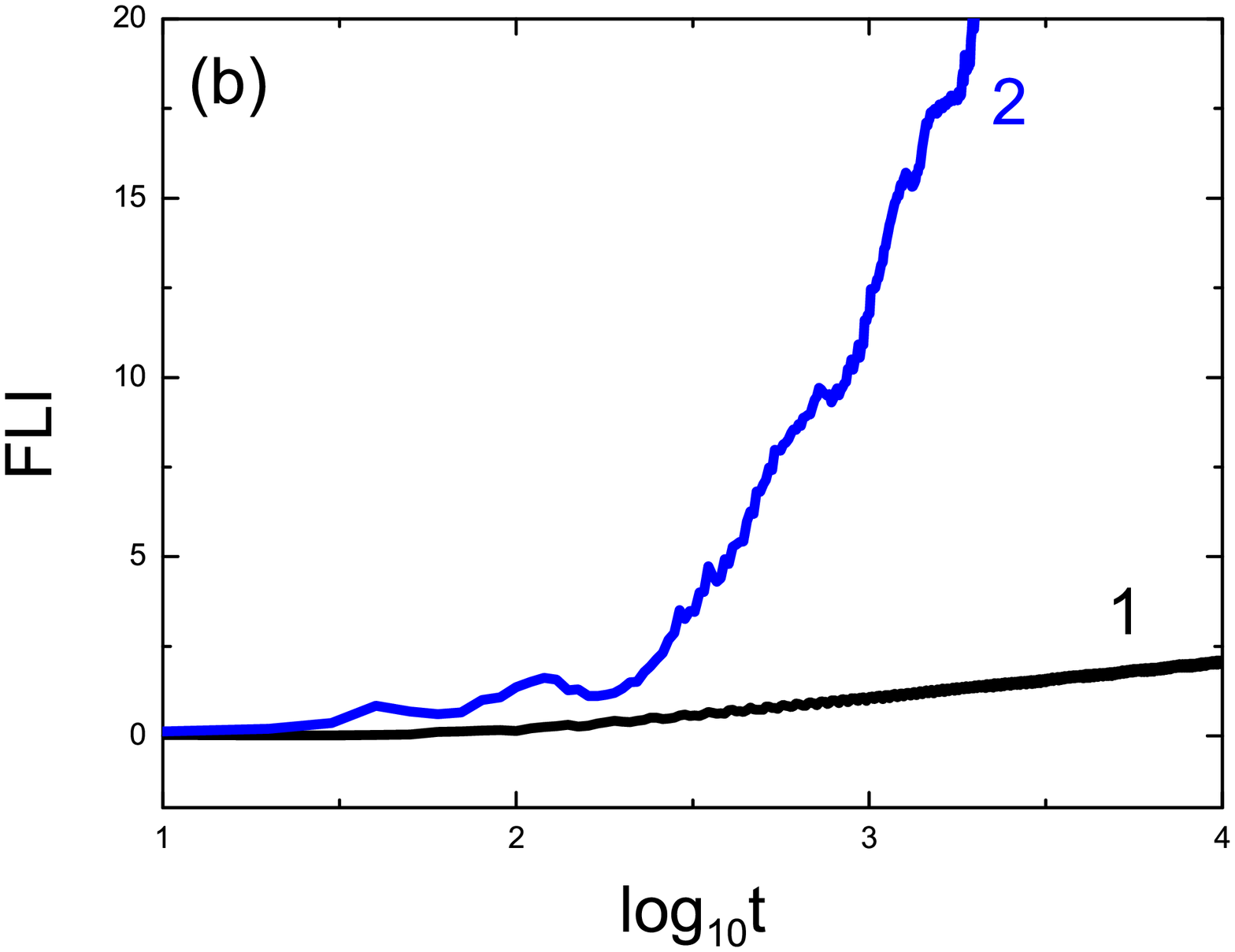}
\caption{Lyapunov exponents $\lambda$ and fast Lyapunov
indicators $\Lambda$ for the new algorithm EC solving orbits 1 and
2 in the FPU-$\beta$ system. Orbit 1 is ordered, whereas orbit 2
is chaotic. }} \label{fig2}
\end{figure*}

\begin{figure*}
\center{
\includegraphics[scale=0.4]{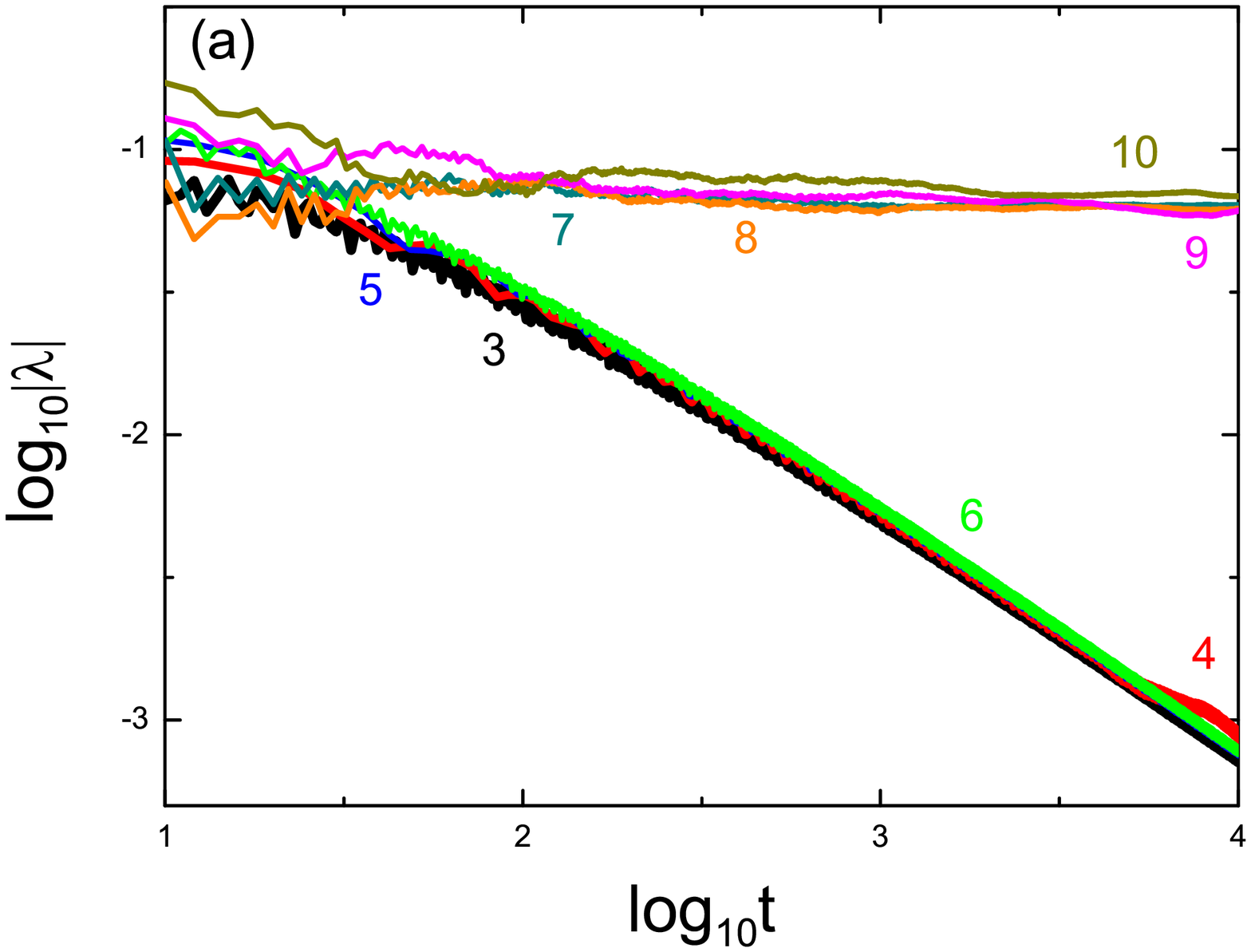}
\includegraphics[scale=0.4]{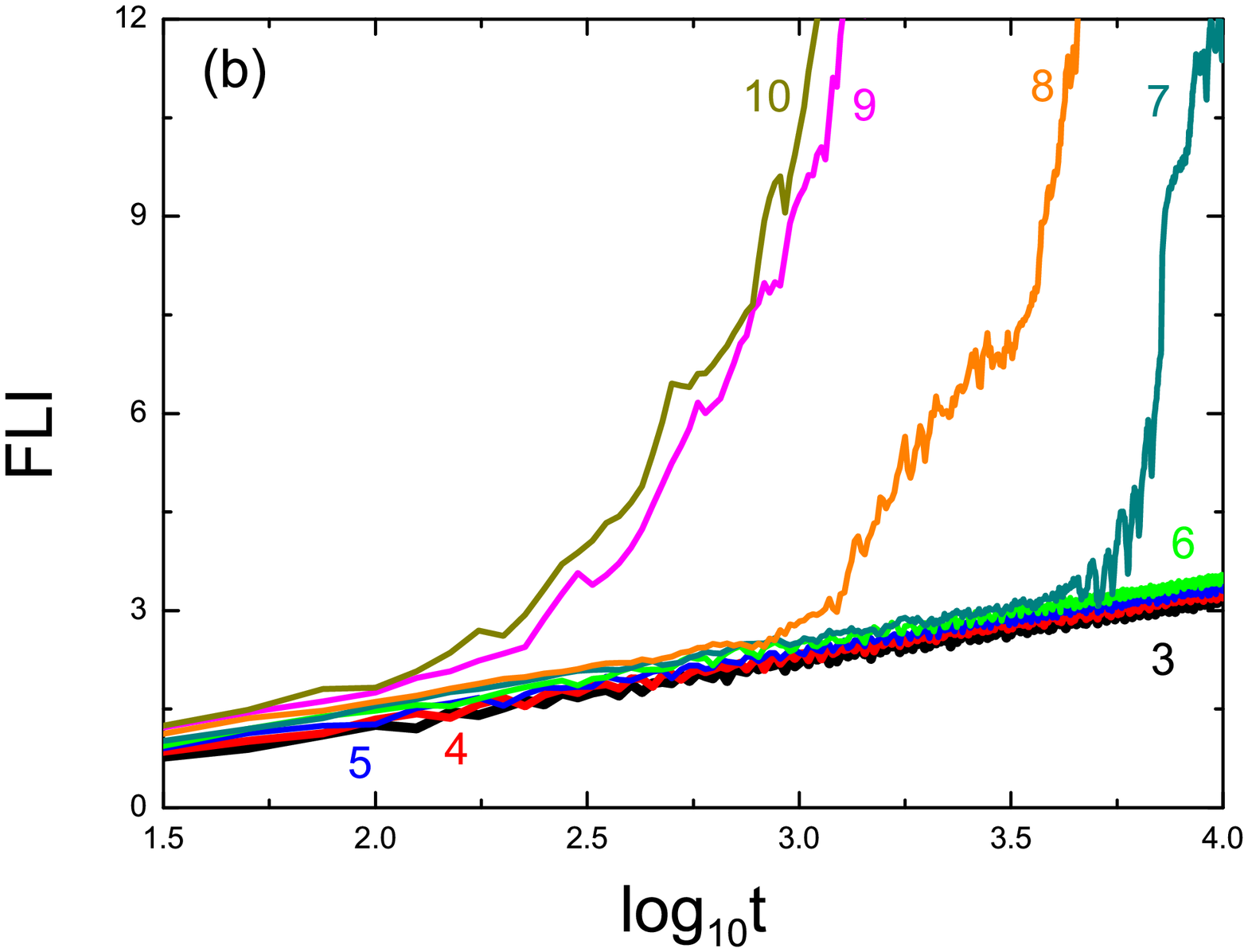}
\caption{Lyapunov exponents $\lambda$ and fast Lyapunov
indicators $\Lambda$ for the new algorithm EC solving other orbits
in the FPU-$\beta$ system. }} \label{fig3}
\end{figure*}

\begin{figure*}
\center{
\includegraphics[scale=0.35]{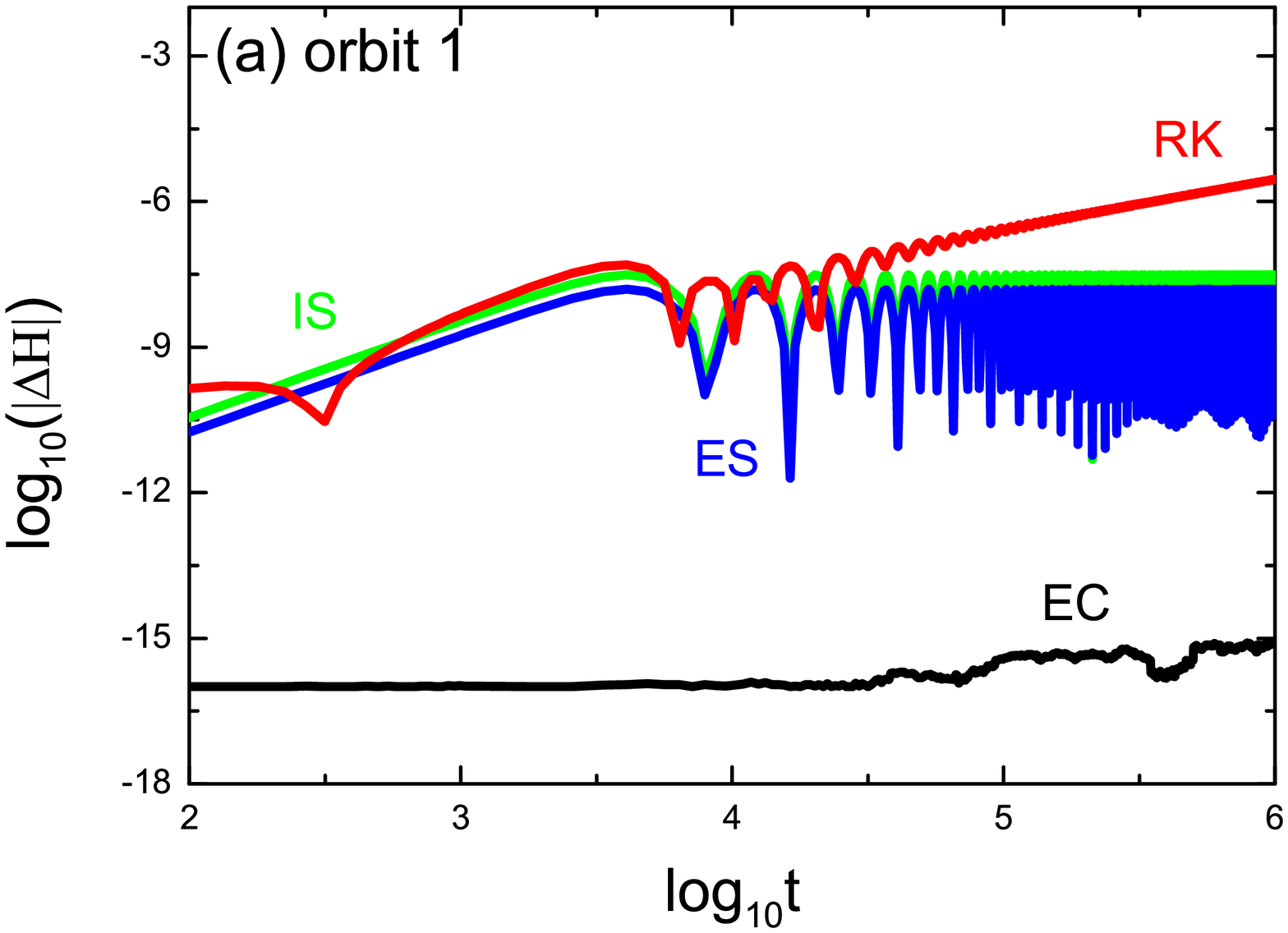}
\includegraphics[scale=0.34]{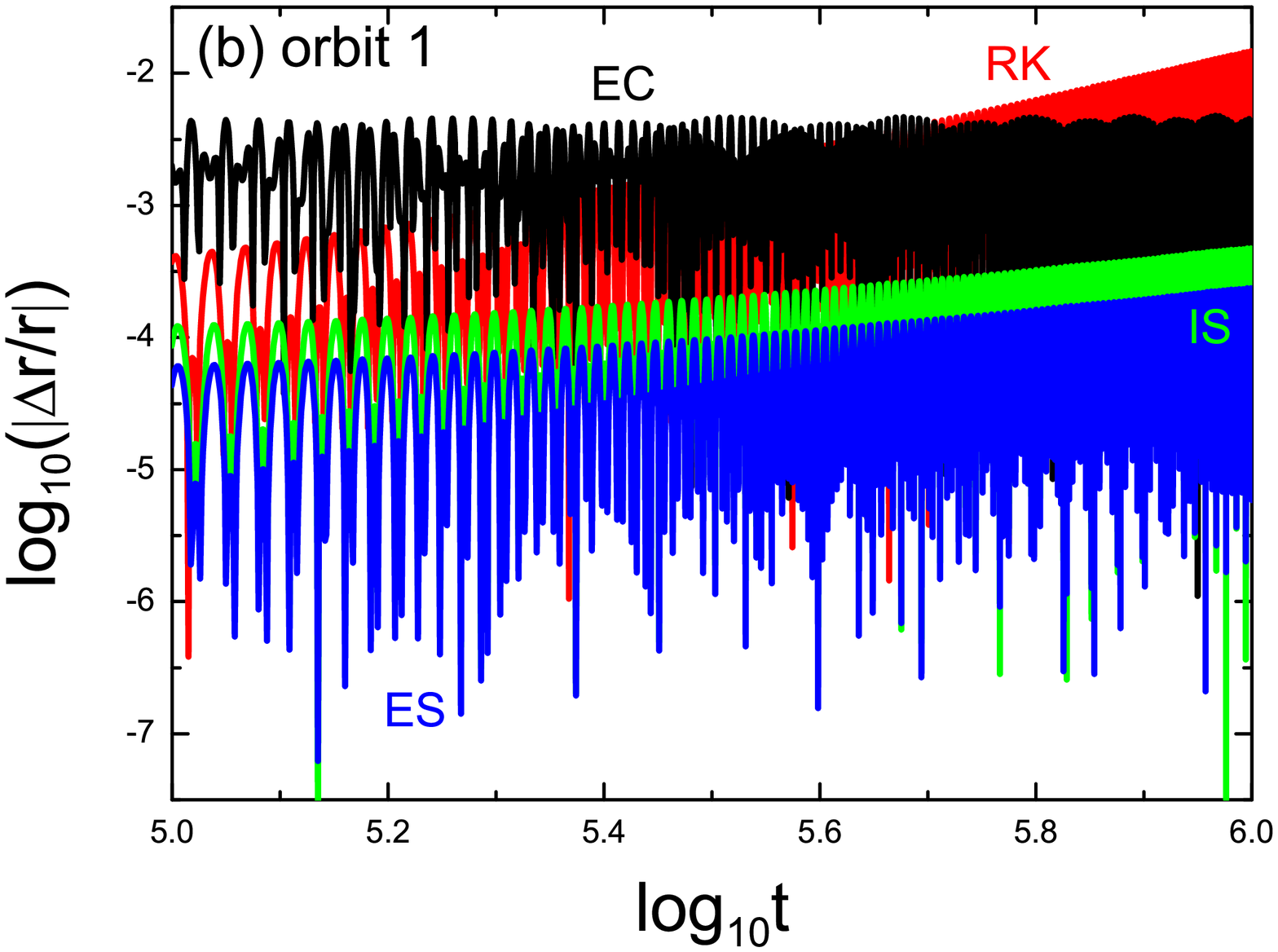}
\caption{Hamiltonian errors, $\Delta H$ and relative position
errors of the four algorithms, EC, RK, ES, and IS solving orbit 1 in
the post-Newtonian problem of compact binaries with one body
spinning, given by Equation (29). The spin-orbit couplings and
spin-spin effects are included together. The initial conditions
and parameters of orbit 1 are given  in Table 2. The time step is
$h=15$. }} \label{fig4}
\end{figure*}

\begin{figure*}
\center{
\includegraphics[scale=0.35]{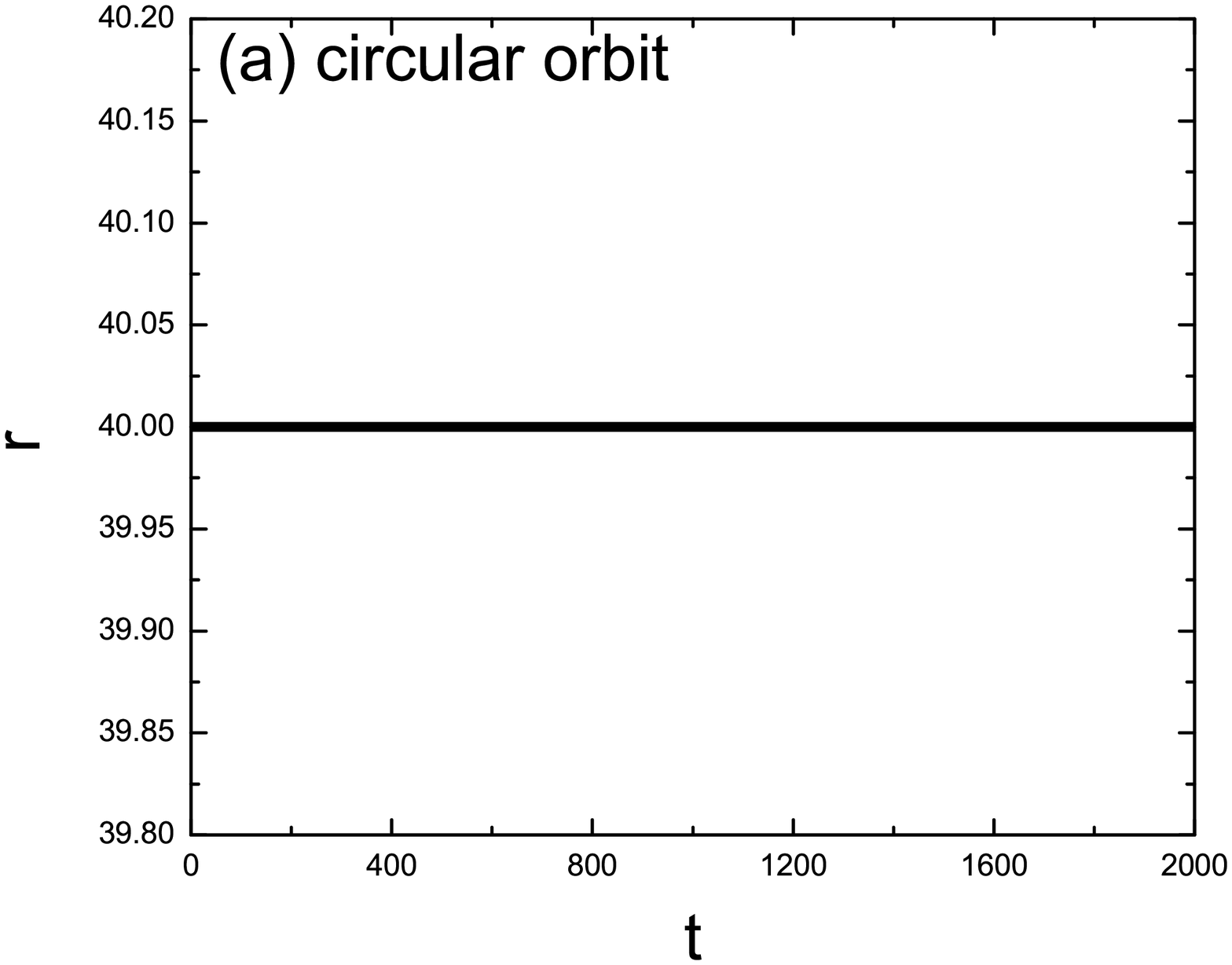}
\includegraphics[scale=0.358]{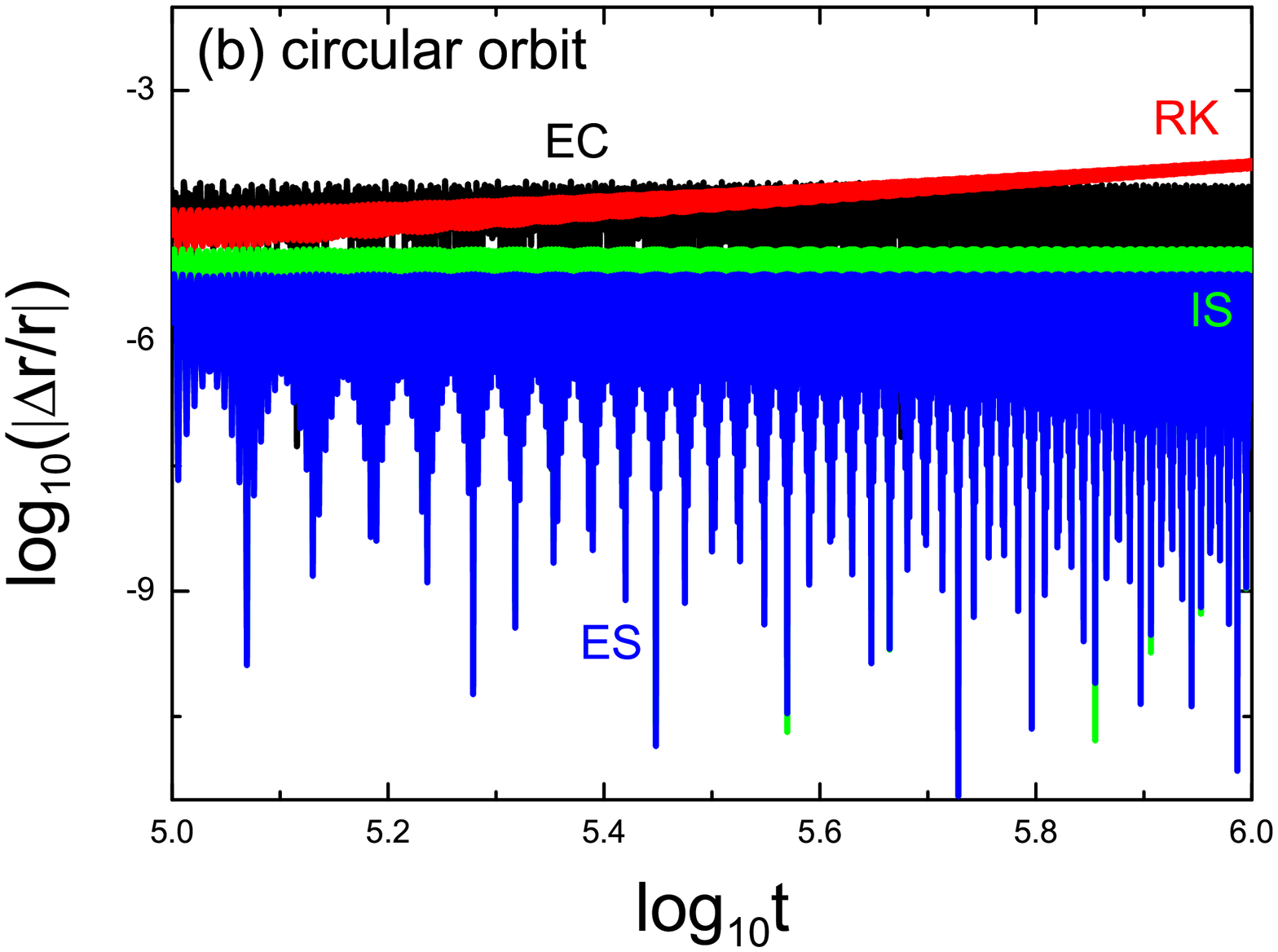}
\includegraphics[scale=0.35]{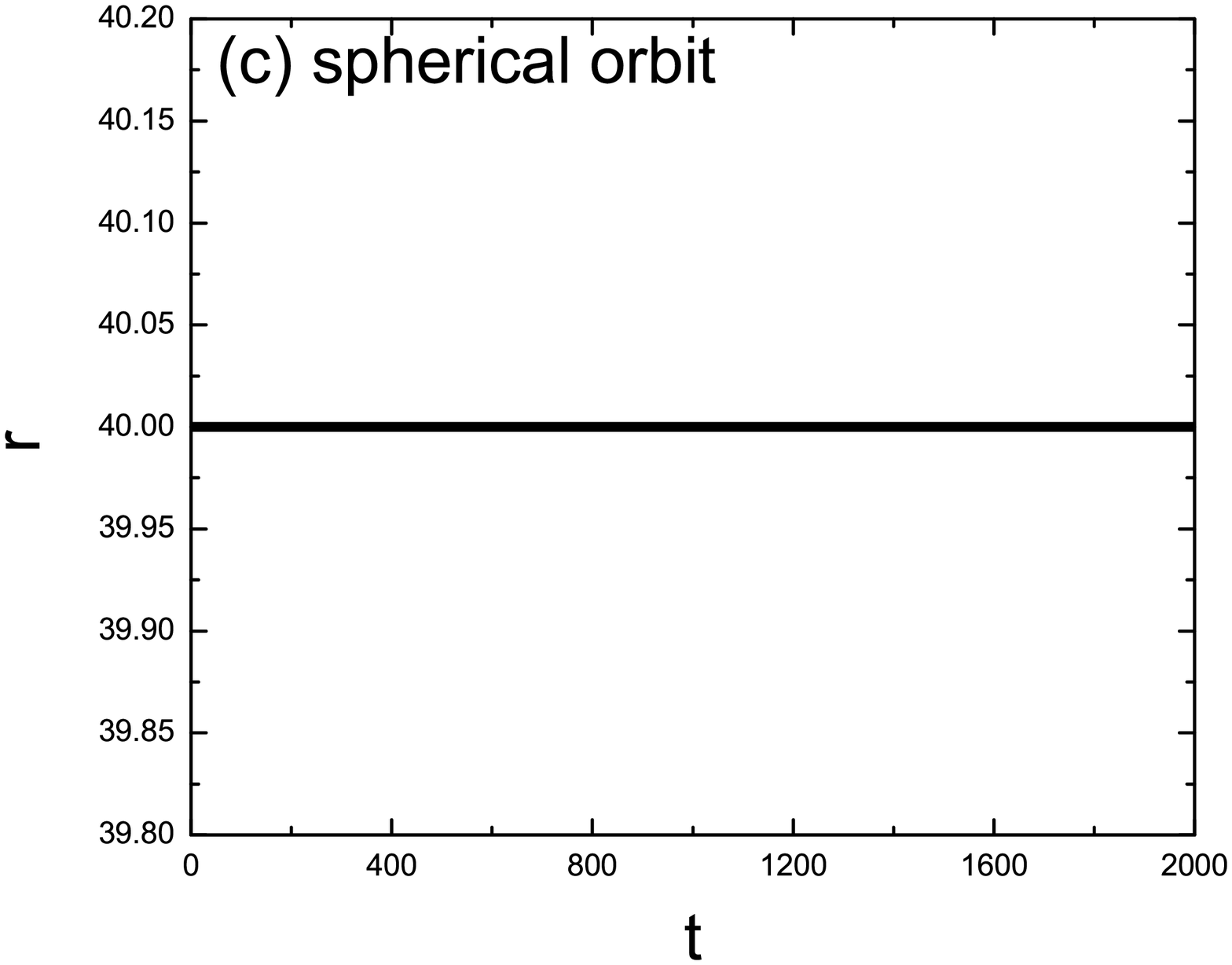}
\includegraphics[scale=0.358]{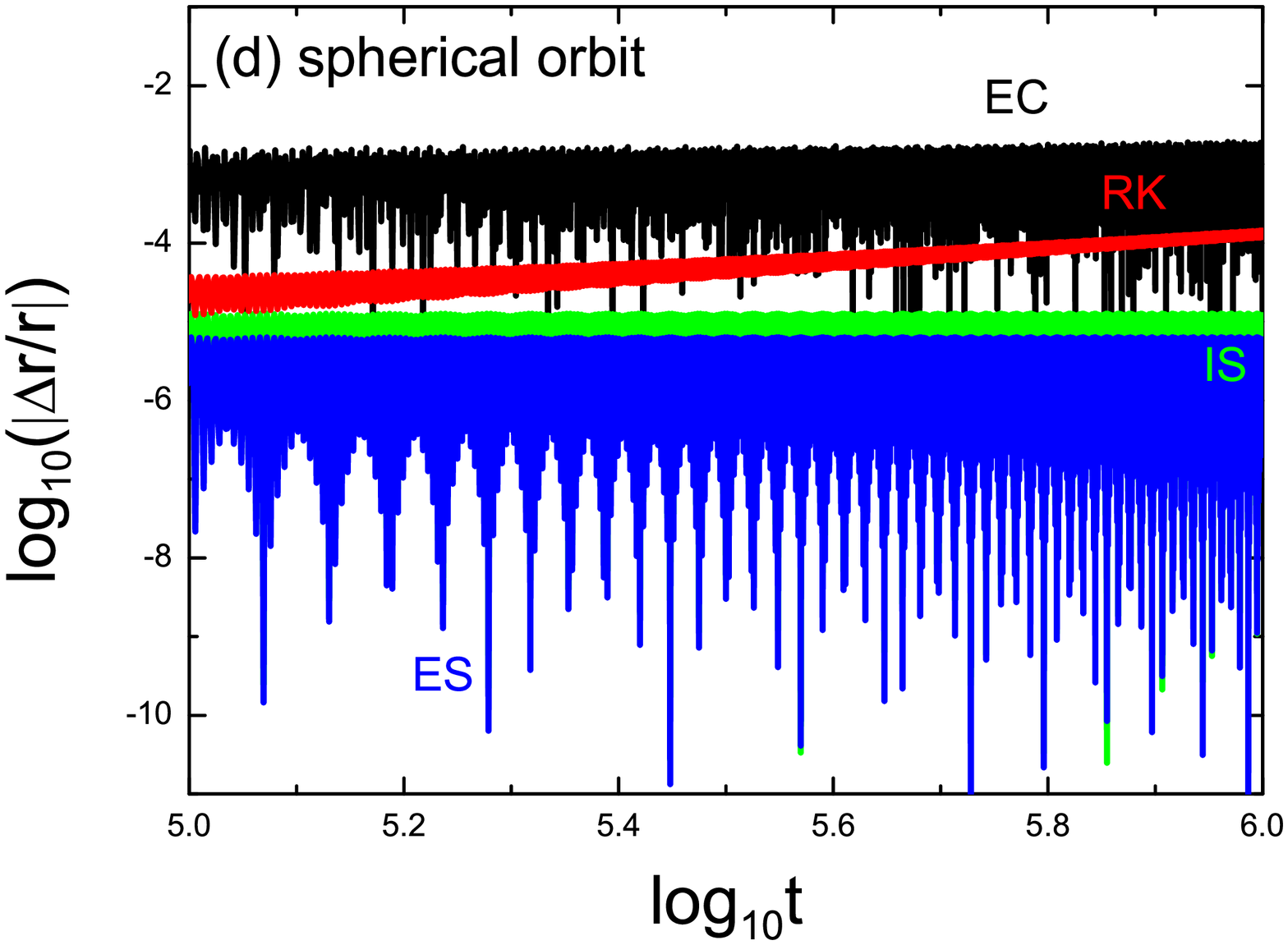}
\includegraphics[scale=0.35]{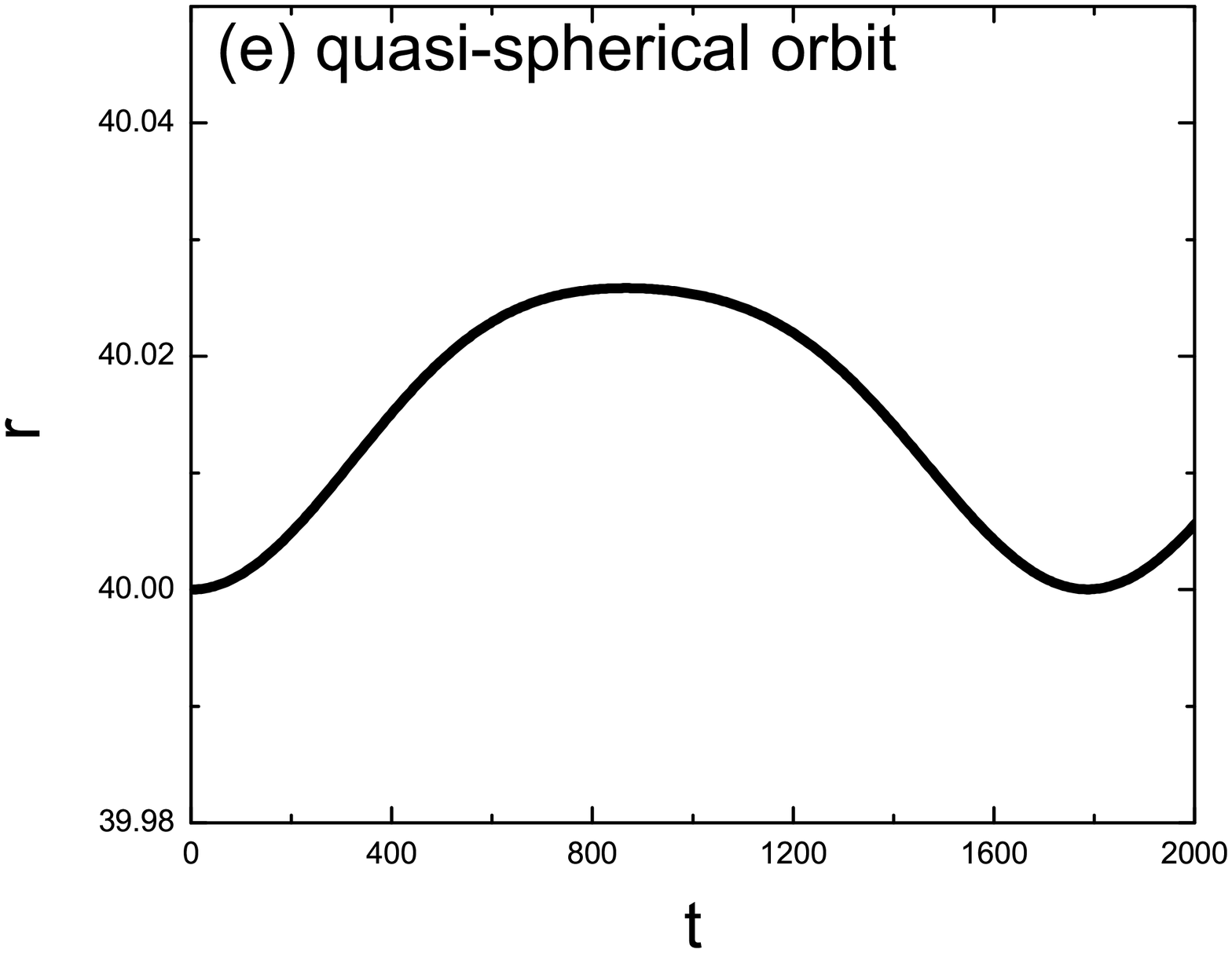}
\includegraphics[scale=0.358]{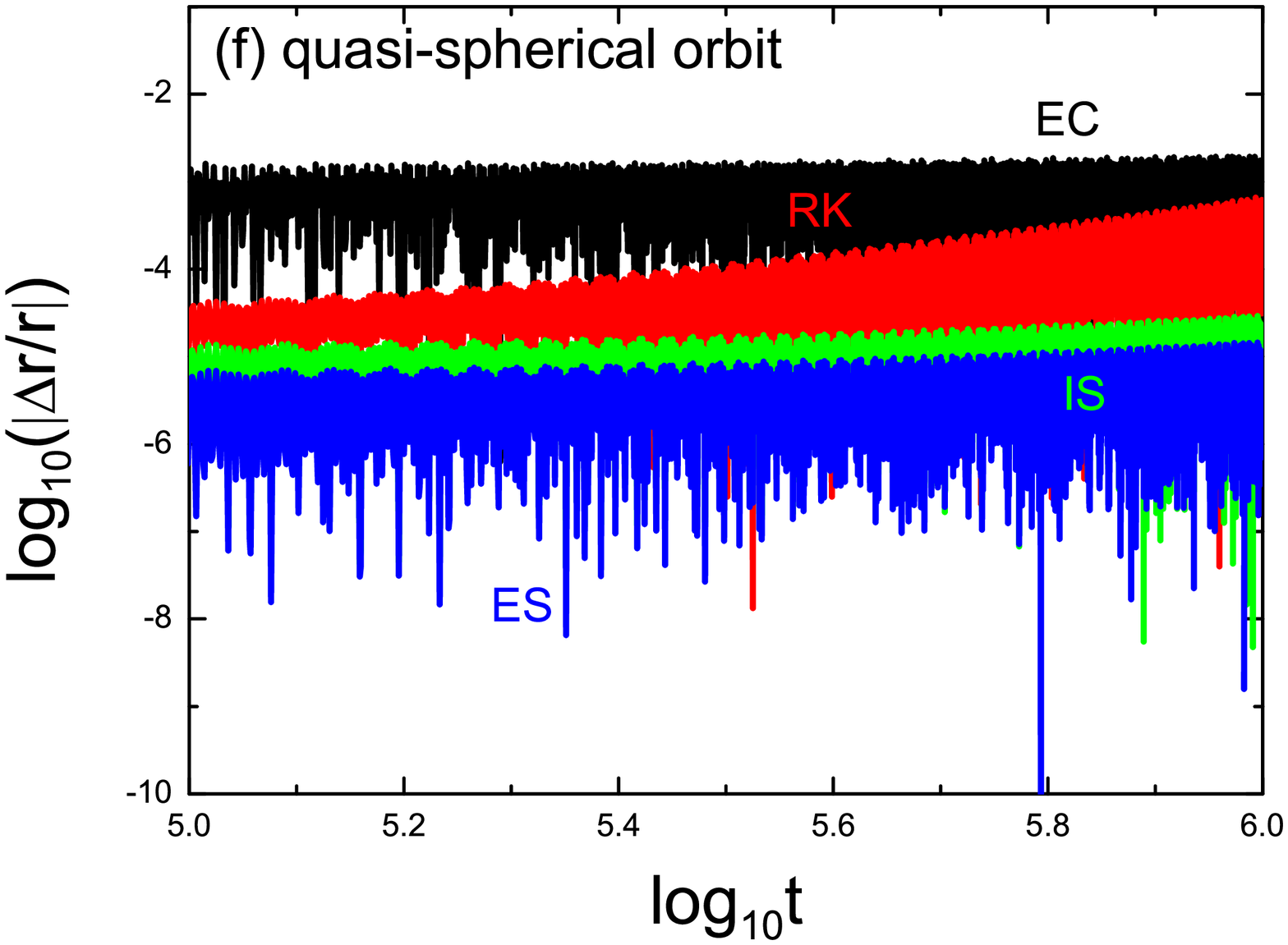}
\caption{Left: evolution of the radii $r$ with time $t$. Right:
relative position errors. Panels (a) and (b): circular orbit in
the non-spinning case. Panels (c) and (d): spherical orbit in the
spin-orbit case. Panels (e) and (f): quasi-spherical orbit with
the inclusion of spin-orbit couplings and spin-spin effects. EC in
(a) and (b) denotes  the energy-conserving method of Bacchini et
al. (2018a) because this circular orbit is considered in a
six-dimensional phase space, but it is our new method in (c)-(f)
because the spherical and quasi-spherical orbits are considered in
a eight-dimensional phase space. The time step is $h=1$. }}
\label{fig5}
\end{figure*}

\begin{figure*}
\center{
\includegraphics[scale=0.5]{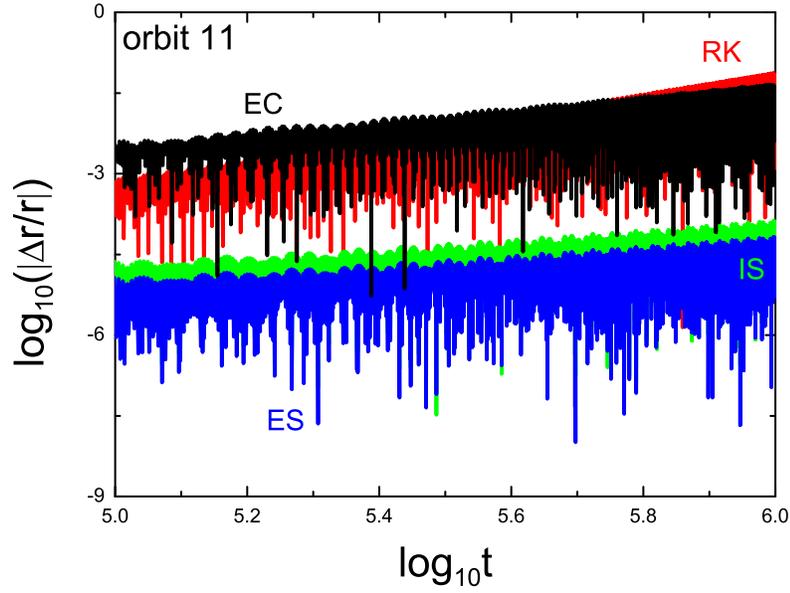}
\caption{Relative position errors for the four algorithms solving
the eccentrical orbit 11 in Table 3. The time step is $h=1$.}}
\label{fig6}
\end{figure*}

\end{document}